\documentclass[useAMS,usenatbib]{mn2e}

\usepackage{graphics}
\usepackage{epsfig}
\usepackage{latexsym}

\usepackage{amsfonts}
\usepackage{amsmath}
\usepackage{amssymb}
\usepackage[english]{babel}
\usepackage{graphicx}
\usepackage{url}
\usepackage{textcomp}

\usepackage{amssymb}
\usepackage{graphics}
\usepackage{graphicx}
\usepackage{epstopdf}

\usepackage[usenames]{color}

\newcommand{\be}{\begin{equation}}
\newcommand{\ee}{\end{equation}}
\newcommand{\ba}{\begin{eqnarray}}
\newcommand{\ea}{\end{eqnarray}}
\newcommand{\bc}{\begin{center}}
\newcommand{\ec}{\end{center}}

\def\mnras{{MNRAS}}
\def\apJ{{ApJ}}
\def\aJ{{AJ}}
\def\apJL{{ApJL}}
\def\aap{{A\&A}}

\def\nat{Nature}

\def\aas{A\&AS}
\def\pasa{{PASA}}

\title[SN 2011fu]{SN~2011fu: A type IIb Supernova with a luminous double-peaked light curve \\}
\author[Morales-Garoffolo, A.]{A. Morales-Garoffolo$^{1}$ \thanks{amg@ice.csic.es}, N. Elias-Rosa$^{2}$,  M. Bersten$^{3}$$^{,}$$^{4}$$^{,}$$^{5}$, A. Jerkstrand$^{6}$,\newauthor S. Taubenberger$^{7}$$^{,}$$^{8}$, S. Benetti$^{2}$, E. Cappellaro$^{2}$, R. Kotak$^{6}$, A. Pastorello$^{2}$, \newauthor F. Bufano$^{9}$,  R.~M. Dom\'inguez$^{10}$, M. Ergon$^{11}$, M. Fraser$^{12}$,  X. Gao$^{13}$,\newauthor E. Garc\'ia$^{14}$$^{,}$$^{15}$$^{,}$$^{16}$, D.~A. Howell$^{17}$$^{,}$$^{18}$, J. Isern$^{1}$, S.~J. Smartt$^{6}$,  L. Tomasella$^{2}$, S. Valenti$^{17}$$^{,}$$^{18}$\\
$^{1}$Institut de Ci\`encies de l'Espai (CSIC-IEEC), Campus UAB,  Cam\'i de Can Magrans S/N, 08193 Cerdanyola (Barcelona), Spain \\
$^{2}$INAF Osservatorio Astronomico di Padova, Vicolo dell'Osservatorio 5, 35122 Padova, Italy\\
$^{3}$Instituto de Astrof\'isica La Plata, IALP (CCT La Plata), CON
ICET-UNLP, Paseo del Bosque s/n, 1900 La Plata, Argentina\\
$^{4}$Facultad de Ciencias Astron\'omicas y Geof\'isicas, Universidad Nacional de La Plata, Paseo del Bosque s/n, 1900 La Plata, Argentina\\
$^{5}$Kavli Institute for the Physics and Mathematics of the Universe (WPI), The University of Tokyo, Kashiwa, Chiba 277-8583, Japan\\
$^{6}$Astrophysics Research Centre, School of Mathematics and Physics, Queen's University Belfast, Belfast BT7 1NN, UK\\
$^{7}$Max-Planck-Institut f\"{u}r Astrophysik, Karl-Schwarzschild-Str. 1, 85741 Garching bei M\"{u}nchen, Germany\\
$^{8}$ European Southern Observatory, Karl-Schwarzschid-Str 2, 85748 Garching bei M\"{u}nchen, Germany\\
$^{9}$INAF-Osservatorio Astrofisico di Catania, Via S.Sofia 78, 95123, Catania, Italy\\
$^{10}$ Instituto de F\'isica de Cantabria (IFCA), Observatorio Astron\'omico de Cantabria, Av. de los Castros, 39005 Santander (Cantabria), Spain\\
$^{11}$The Oskar Klein Centre, Department of Astronomy, Stockhom University, Albanova, 10691 Stockholm, Sweden\\
$^{12}$Institute of Astronomy, University of Cambridge, Madingley Road, Cambridge CB3 0HA, United Kingdom\\
$^{13}$No.1 Senior High School,Urumqi, Xinjiang,China\\
$^{14}$Departamento de F\'isica Aplicada I, E.T.S. Ingenier\'ia, Universidad del Pa\'is Vasco, 9 Alameda Urquijo s/n, 48013 Bilbao, Spain\\
$^{15}$Esteve Duran Observatory, Foundation Observatori Esteve Duran. Avda. Montseny 46, Seva 08553, Spain\\
$^{16}$Institut d'Estudis Espacials de Catalunya (IEEC), Edif. Nexus, C/ Gran Capit\`a 2-4, 08034 Barcelona, Spain\\
$^{17}$Las Cumbres Observatory Global Telescope Network, 6740 Cortona Dr., Suite 102, Goleta, CA 93117, USA\\
$^{18}$Department of Physics, University of California, Santa Barbara, Broida Hall, Mail Code 9530, Santa Barbara, CA 93196-9530, USA\\
}

\begin{document}

\date{}

\pagerange{\pageref{firstpage}--\pageref{lastpage}} 
\pubyear{2015}

\maketitle

\label{firstpage}

\begin{abstract}
We present optical and near infrared observations of the type IIb supernova (SN)~2011fu from a few days to $\sim300$~d after explosion. The SN presents a double-peaked light curve (LC) similar to that of SN~1993J, although more luminous  and with a longer cooling phase after the primary peak. The spectral evolution is also similar to SN~1993J's, with hydrogen dominating the spectra to $\sim40$~d, then helium gaining strength, and  nebular emission lines appearing from $\sim60$~d post-explosion.  The velocities derived from the P-Cygni absorptions are overall similar to those of other type IIb SNe.  We have found a strong similarity between the oxygen and magnesium line profiles at late times, which suggests that these lines are  forming at the same location within the ejecta.  The hydrodynamical modelling of the pseudo-bolometric LC and the observed photospheric velocities suggest that SN~2011fu was the explosion of an extended star ($\rm R\sim450$ R$_\odot$), in which  1.3 $\times 10^{51}$ erg of 
kinetic 
energy were 
released and 0.15~M$_{\rm \odot}$ of $^{56}$Ni were synthesised. In addition, a better reproduction of the observed  early pseudo-bolometric LC is achieved if a more massive H-rich-envelope than for other type IIb SNe is considered (0.3~M$_{\rm \odot}$). The hydrodynamical modelling of the LC and the comparison of our late-time spectra with nebular spectral models for type IIb SNe, point to a progenitor for SN~2011fu with a ZAMS mass of 13-18~M$_{\rm \odot}$.

\end{abstract}


\section{Introduction}

Core-Collapse-Supernovae (CC-SNe) are believed to be the explosions that mark the end-point in the evolution of massive stars \citep[$M_{\rm ZAMS} > 8$ M$_{\odot}$; see e.g.][]{heger03}. They are spectroscopically divided in different groups according mainly to the presence of the H and He lines in their optical spectra.  While type II SNe show prominent H features, type I do not, and can be further subdivided, according to the presence or absence of He, as types Ib and Ic respectively. There are also hybrid objects which undergo a transition from being H dominated at early phases of their spectral evolution, to He dominated at later times. These are termed type IIb SNe. The first SN to have shown this spectral transition was SN~1987K,  and its progenitor was proposed to be a star that had lost most of its hydrogen envelope before exploding \citep{Fpk88}. The best studied type IIb SN to date is SN~1993J (e.g. \citealt{Nomoto93J,Podsiadlowski93J,Ws94}), that is considered the prototype 
of the 
subclass.
 Its LC showed an early peak, followed by a rapid decline thought to be the consequence of the cooling of the  progenitor's stellar envelope after shock breakout. Subsequently it presented a secondary maximum attributed to input from the radioactive decay of $^{56}$Ni. The star that exploded as SN~1993J was detected in archival images \citep{Al94}, and its disappearance was confirmed by \cite{Mnd09}. 
 Two mechanisms have been proposed by which type IIb SN progenitors lose part of their H envelope before exploding, namely stripping by a close companion after Roche Lobe overflow, and wind-driven stellar mass loss in a single star. The former scenario has gained strength over the past years, for example with the detection of signatures of the possible hot blue companion star of SN~1993J (\citealt{Mnd04}; see also \citealt{Fox14}), and the modelling of a progenitor binary system and likely detection of the stellar companion of SN~2011dh \citep{Benvenuto13,Folatelli14}. Moreover, although a Wolf-Rayet (WR) wind-like spectrum was obtained for the type IIb SN~2013cu a few hours after explosion \citep{Gal-Yam14}, recent work has shown that the progenitor star for SN~2013cu was unlikely a WR \citep{groh14,shivvers14,smith15}.

 An important question is which of the observed properties of SNe IIb can be connected with the characteristics of their progenitor systems. For example, \cite{ecIIb} proposed that bright early luminosity from the shock-heated progenitor stellar envelope, low radio shell velocities, and 
thermal X-ray emission were characteristic of extended progenitors with R~$\sim 150$~R$_\odot$, while compact progenitors with R~$\sim 1$~R$_\odot$ have faint early optical LCs, have high radio shell velocities and non-thermal X-ray emission. However, this link is not straightforward  and needs to be revised since the type IIb SNe~2011dh and 2011hs present fast expanding radio shells but the hydrodynamical modelling of their LC \citep{Melina12,bufano14}, and in the case of SN~2011dh the detection of its progenitor in archival images \citep{Mnd11,VD11} and its disappearance \citep{VD13}, point to the explosion of extended stars. To further complicate the picture, SNe IIb show a variety of continuum flux excess in early UV spectra \citep{Ben-Ami14}. Strong UV-excess suggest significant interaction between the SN ejecta and circumstellar material. 

To date, few type IIb SNe have been thoroughly studied since they are relatively rare events. \cite{Li11} estimated the fraction of type IIb over a volume limited sample of 81 type II SNe to be $11.9\% ^{+3.9}_{-3.6}$. In this paper we present optical and near infrared (NIR) data for the type IIb  SN~2011fu.  SN 2011fu, with coordinates $\alpha =02^{\rmn{h}} 08^{\rmn{m}} 21.26\fs $ and $\delta = 41\degr 29\arcmin 9.9\arcsec$ (J2000), was discovered in a spiral arm of the galaxy UGC~1626 by F. Ciabattari and E. Mazzoni of the Italian Supernovae Search Project (ISSP)\footnote{http://italiansupernovae.org/en.html}, on 2011 September 21.04 UT and classified by \cite{tomasella11} as a young type II SN 2011 September 23.84 UT.
A previous analysis of optical data of SN~2011fu was presented by \cite{11fu}, which confirmed it was a type IIb SN.
In this manuscript we contribute with data probing the whole SN evolution. In Section \ref{host} we discuss the distance, reddening, and explosion epoch of SN~2011fu. In Section \ref{data} we summarize our observations and the reduction process of the data. In Section \ref{phot} we present the optical and NIR LCs of the object, while in Section \ref{spec} we present and discuss the spectral evolution. In Section \ref{discussion} we discuss the results obtained from the hydrodynamical modelling of the pseduo-bolometric LC and we compare our nebular spectra with published models for late time type IIb SN spectra. Finally, in Section \ref{conclusion}, we present the conclusions of our work.

\section{Distance, Reddening, and explosion epoch}
\label{host}

The rise to a first maximum in the \textit{V} and \textit{R} LCs (see the inset of Figure \ref{fig:apparentLCs}) implies that SN~2011fu was discovered very soon after explosion. In fact, models show that the evolution of the SN LC during this rise  should be very fast, and lasts $\sim 1$~d for a number of configurations (e.g. \citealt{1998ApJ...496..454B,Melina12}). By adopting  $\rm JD_{\rm explo}=2455824.5 \pm 0.7$ as the explosion date of SN~2011fu, we obtain the best fit of the early phase of the pseudo-bolometric LC with the hydrodynamical models presented in Section \ref{LCmodelling}. For this reason we have chosen $\rm JD_{\rm explo}=2455824.5 \pm 0.7$ as the explosion date of SN~2011fu, which is in agreement with the  discovery epoch, the last non detection of the object which  was 2011 August 10 ($\rm JD=2455783.5$), the pre-discovery image taken by Xing Gao 2011 September 20 ($\rm JD=2455825.2$), and the classification as a young object. Note that this assumed explosion date 
also agrees, within the uncertainties, with the explosion epoch adopted by \cite{
11fu}. All phases in the rest of this manuscript are given with respect to $\rm JD_{\rm explo}=2455824.5 \pm 0.7$.

UGC~1626 is a SAB(rs)c type galaxy whose recessional velocity given by the NASA/IPAC extragalactic database (NED) is $5543 \pm 11$~km~s$^{-1}$. We have performed measurements of the SN redshift from the narrow H$\alpha$ emissions in its spectra and have obtained the same redshift as for its 
host galaxy, which we adopt in the rest of this paper. The redshift derived distance to UGC~1626, also provided by NED assuming H$_{0}=73\pm5$ km s$^{-1}$ Mpc$^{-1}$, $\Omega_{\rm m}=0.27$, $\Omega_{\rm v}=0.73$, and accounting for the Virgo, Great Attractor and Shapley infalls  is  $74.5 \pm 5.2$ Mpc, i.e. $\rm \mu = 34.36 \pm 0.15 $ mag, \citep{Mould00}.

The reddening in the line of sight of UGC~1626 due to the Milky Way is $E(B-V)_{{\rm MW}}=0.068\pm0.002$~mag \citep{Schfink}.  We detect a narrow absorption likely due Na {\sc i} D $\lambda\lambda$$5890$, $5896$ at the host galaxy redshift. The equivalent width of this line is $0.33\pm0.03$~\AA{} that using the relations for unresolved Na {\sc i} D given by \cite{Poznanski12}, provides an estimate of the reddening in the host galaxy as: $E(B-V)_{{\rm host}}=0.035^{+0.042}_{-0.029}$~mag. In fact, a similar value of reddening due to the host galaxy is obtained using the relations of \cite{Turatto03}.  In the rest of this manuscript we adopt $E(B-V)_{{\rm total}}=0.10^{+0.04}_{-0.03}
$~mag as the total reddening towards SN~2011fu. We note that in \cite{11fu} the relations of \cite{MunariZwitter97} were used to 
estimate the contribution to the reddening due to the host galaxy, and they obtained $E(B-V)_{{\rm total}}=0.22\pm0.11$, which is consistent with the value we have adopted within the uncertainties.

\section{Observational data, reduction, and calibration process}
\label{data}
The observational follow-up data of SN 2011fu presented in this paper covers the period between a few days post discovery, until the SN disappeared behind the Sun (end of February 2012). After that, one more spectrum and one epoch of \textit{gri} photometric data were obtained  well into the nebular phase of the SN, 2012 July 20 UT. 
Thanks to a large collaboration between many European institutions, optical  photometry and spectroscopy, as well as some NIR photometric data of SN 2011fu, have been collected at several sites. 
Some amateur data taken by F.Ciabattari and E. Mazzoni from Italy, and Xing Gao from China were also included in our analysis. A summary of the characteristics of the instruments used to acquire our photometric data can be seen in Table \ref{photometric_log}.

The list with our spectroscopic observations and the instrumental configuration used to acquire the data is presented in Table \ref{spectroscopic_log}.

\begin{figure}
 \includegraphics[width=8.9cm]{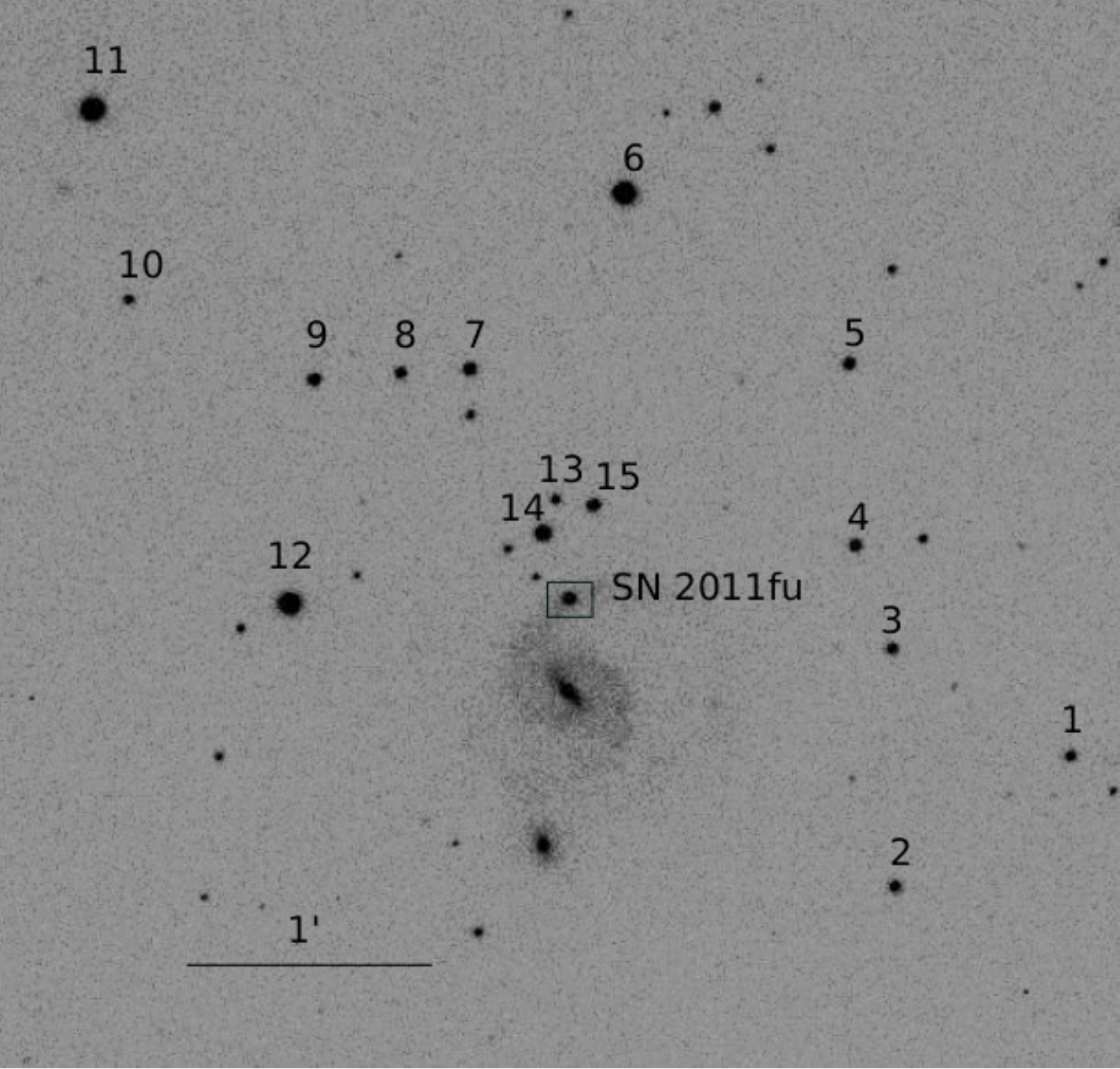}
 \caption{\textit{V} band image of UGC~1626 taken with the 2.2m Calar Alto Telescope + CAFOS on 2011 October 16. The stars used for the photometric calibration of SN~2011fu are labelled. North is to the top and East to the left. }
 \label{fig:11fusequence}
\end{figure}

All the data  were corrected for overscan, bias, and  flat-field within the {\sc iraf}\footnote{Image reduction and Analysis Facility, a software system distributed by the National Optical Astronomy Observatories (NOAO)} environment, except for data obtained at the Liverpool Telscope (LT) which were automatically reduced with the instrument specific pipeline. The instrumental magnitudes of the SN were derived via Point Spread Function (PSF) fitting which was done with {\sc snoopy} \citep{cappellaro14}. In the case of the \textit{U} and \textit{u} photometry, the measurements were done after performing template subtraction. For all other bands, PSF photometry at late time was verified by performing template subtractions and no major differences were found\footnote{The \textit{UBVRI} band templates were obtained on 2012 October 23 with AFOSC at the 1.82 m telescope in the 
Asiago Observatory, while the \textit{uri} band templates were obtained on 2014 October 25 at the LT with IO:O in the \textit {Roque de los Muchachos} Observatory.}.  A set of fifteen local stars in the SN field (Figure \ref{fig:11fusequence})  were used to trace the photometric calibration. The stellar sequence was calibrated with the zero points and colour terms derived  for each site thanks to observations of 
Landolt standards (Tables \ref{sequence_magsJC} and \ref{sequence_magssloan}) and
the SN magnitudes were calculated relative to 
these. The errors of the magnitudes of the stellar sequence were obtained as the rms deviation with respect to their mean values. Given the fact that several instruments with different passbands were used to acquire our data, we performed S-corrections of our \textit{BVrRiI} photometry to the Bessell system \citep{stritz02,pignata04}. In order to derive the S-correction terms we used our spectral sequence of the SN. Note that we were not able to perform S-corrections to the \textit{U}, \textit{u}, and 
\textit{z} band 
photometry given the fact that our spectra do not cover completely these passbands. The SN magnitudes together with their errors, calculated by adding in quadrature the uncertainties associated with the calibration and the PSF fit\footnote{Obtained by placing artificial stars with magnitudes similar to those of the SN at positions near the SN and calculating the standard deviation of their recovered magnitudes.}, are reported in Tables \ref{UBVRI_phot}, \ref{ugz_phot}, and \ref{jhk_phot}.

The unfiltered photometry provided by F. Ciabattari was rescaled to the \textit{R} band considering that the detector sensitivity peaks at this wavelength. With the same argument the data provided by X. Gao were rescaled to \textit{V}. 

 Note that stars 2, 6, 9, 12, and 14 from our sequence coincide with  stars 2, 1, 4, 3 and 5 used for photometric calibration by \cite{11fu}. The greatest difference between the two sequences is 0.1 mag found for star 2 in the \textit{U} band. For the rest of the bands differences are found to be less than 0.1 mag.  

The NIR data was reduced within {\sc iraf} with the help of {\sc caindr}, a package developed by Jos\'e A. Acosta Pulido and Rafael Barrena for the instrument CAIN3 at the TCS. The images of each specific filter were corrected for flat field, and  bad pixel corrections were applied. Then background images were constructed and subtracted from the science frames, and finally the images were combined taking into account the dithering pattern. Similarly to the optical data, we measured the instrumental SN magnitudes by PSF fitting. The SN calibration was done relative to 2-MASS stars in the field (\citealt{2mass06}; see Table \ref{sequence_magsJC}).

The spectra were corrected for overscan, bias, and flat-field within {\sc iraf}. Next they were variance-weighted extracted and wavelength calibrated  with the use of arc lamp spectra. The wavelength calibration was cross-checked with respect to sky emission lines. Finally a flux calibration was applied using sensitivity functions obtained from the observation of spectrophotometric standards. The flux calibration was cross-checked with the photometry of the SN at the nearest epoch. Telluric lines were corrected dividing the SN spectra by telluric line spectra obtained from the spectrophotometric standards.

\section{Photometry}
\label{phot}

\subsection{Light curves}
\begin{figure*}
 \includegraphics[scale=0.7]{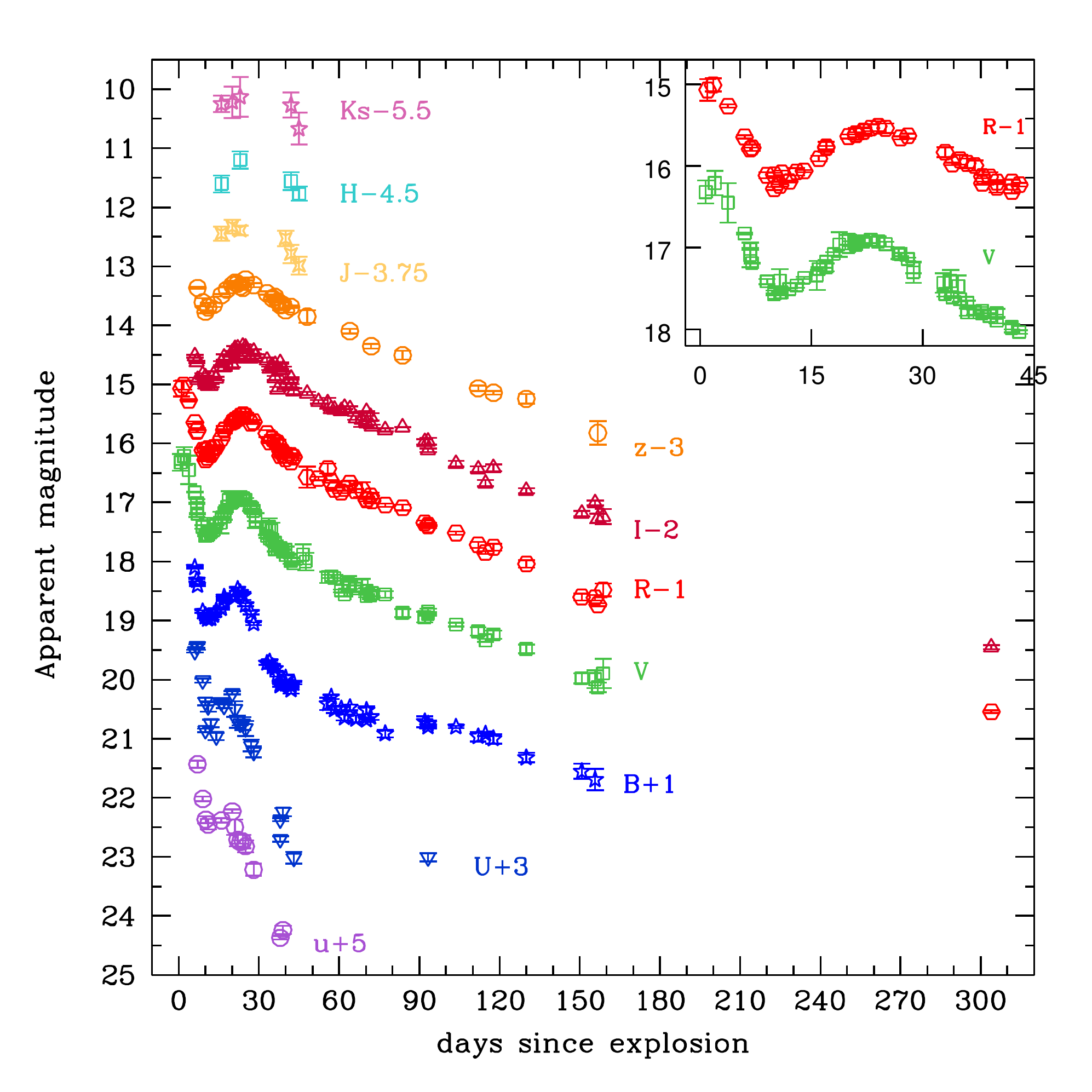}
 \caption{Optical-NIR light curves of SN~2011fu. The assumed explosion epoch is $\rm JD = 2455824.5\pm0.7$. The LCs have been shifted for clarity by the values indicated in the figure. The inset in the top right corner of the figure is a zoom of the V and R LCs up to 45~d. A colour version of this figure can be found in the online journal.}
 \label{fig:apparentLCs}
\end{figure*}

The optical-NIR LCs of SN~2011fu are presented in Figure \ref{fig:apparentLCs}. In addition to the data presented in the figure we obtained two epochs of Sloan \textit{g} band data, which are listed in Table \ref{ugz_phot}. In the optical, the LCs clearly present two peaks in all bands, but not in the NIR since the data do not cover the early phases. The rise to primary peak in the \textit{V} and \textit{R} band can be seen thanks to the early time data provided by amateur astronomers (see inset in the top-right of the figure).  

By making low order polynomial fits to the optical LCs we have estimated the phases and magnitudes at which the minima after first peak and secondary maxima take place. The results are presented in Table \ref{min_max_decay_rise}. Note that we obtain different absolute magnitudes at secondary peak than \cite{11fu} due to the fact that we have adopted different distance and  reddening towards the host, however, our values agree within the uncertainties. We also fitted low order polynomials to the first peaks in the \textit{R} and \textit{V} LCs. In the \textit{V} band the first peak at 16.3 mag is reached $\sim2.3$~d after explosion while in the \textit{R} band the 16.0 mag first peak is obtained $\sim2.8$~d post explosion.  

We have also estimated the decline rates of the tails of the \textit{BVRIz} LCs (Table \ref{min_max_decay_rise}). These rates are quite similar to those estimated for SN~2013df by \cite{13dfMG}, and are steeper than expected from $^{56}$Co decay. This is a common characteristic to stripped envelope SNe (e.g., SN 2008ax, \citealt{tau08ax}) and is possibly due to increasing transparency for $\gamma$ rays in their lower mass ejecta. 

\begin{table*}
\centering
\caption{Optical and NIR magnitudes of the minimum and secondary maximum of SN~2011fu, the corresponding times at which they occurred, and tail decline rates in the \textit{BVRIz} LCs.}
\begin{tabular}{ccccccc} \hline 
    Band    &    $t_{\rm min}^{a}$   &  Apparent magnitude at minimum   & $t_{\rm max}^{a} $                & Apparent  magnitude at  & Absolute magnitude at & Decline tail \\

            & & & &secondary maximum & secondary maximum&\\
            &    (d)                  &    (mag)                             &   (days)                              &   (mag)                             & (mag)& [mag (100d)$^{-1}$]*\\
  \hline
\textit{U}  &$13.6\pm1.1$ &  $17.90\pm0.01$     & $18.8\pm0.8$ & $17.29\pm0.01$  & $-17.58\pm 0.16$ & --   \\
\textit{B}  &$11.8\pm1.0 $ & $17.96\pm0.01$      & $21.8\pm1.6$ & $17.52\pm0.01$  &  $-17.26\pm0.15 $ & $1.25\pm 0.07$ \\ 
\textit{V}  &$10.6\pm0.7$  &  $17.55\pm0.01$     & $22.9\pm0.7$ & $16.92\pm0.01$  & $-17.76\pm0.15 $  & $1.78 \pm 0.04$   \\ 
\textit{R}  &$11.1\pm2.1  $&  $17.18\pm0.01$     & $24.2\pm1.2$ & $16.53\pm0.01$  & $-18.08\pm0.15 $  & $2.04 \pm 0.04$  \\ 
\textit{I}  &$11.4\pm1.5 $ & $16.95\pm0.01$      & $24.1\pm1.2$ & $16.42\pm0.01$  & $-18.12\pm0.15 $  & $1.97 \pm 0.05 $  \\ 
\textit{z}  &$10.5\pm0.7 $ & $17.29\pm0.01$      & $24.1\pm1.0$ & $16.29\pm0.01$  & $-18.22\pm0.15 $  &  $1.82\pm 0.04$ \\ 
\textit{J}  &--              &  --                 & $24.5\pm2.6$ & $16.08\pm0.01$  & $-18.38\pm0.15 $ &  --  \\ 
\textit{H}  &--              &  --                 & $29.6\pm0.7$ & $15.62\pm0.01$  & $-18.80\pm0.15 $ & --   \\ 
\textit{K$_{s}$}  &--        &  --                 & $29.2\pm4.5$ & $15.56\pm0.01$  & $-18.84\pm0.15 $  & --  \\ 
\hline
\end{tabular} 
\begin{flushleft}
  $^{a}$  $t_{\rm min}$ and $t_{\rm max}$ are calculated with respect to our adopted explosion date $\rm JD =2455824.5\pm0.7$. The errors in the NIR are large due to less photometric coverage of the maxima.\\
  *Considering the interval starting $\sim 40$ days after explosion to $\sim160$ days  \\
 
\end{flushleft}

\label{min_max_decay_rise}
\end{table*}

\subsubsection{Colour curves}
\label{colours}

\begin{figure*}
 \includegraphics[width=8cm]{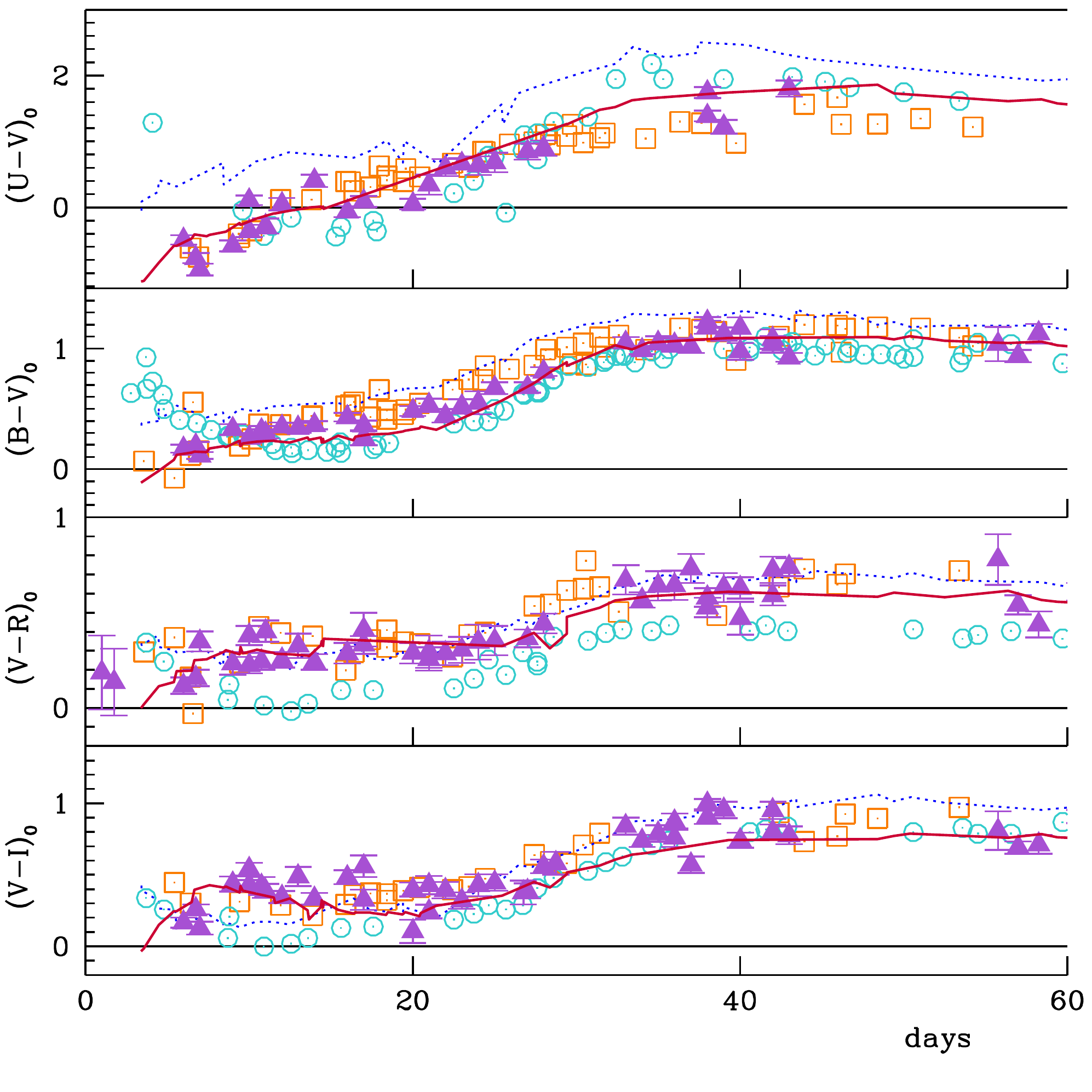}
\hspace{-1cm}
 \includegraphics[width=8cm]{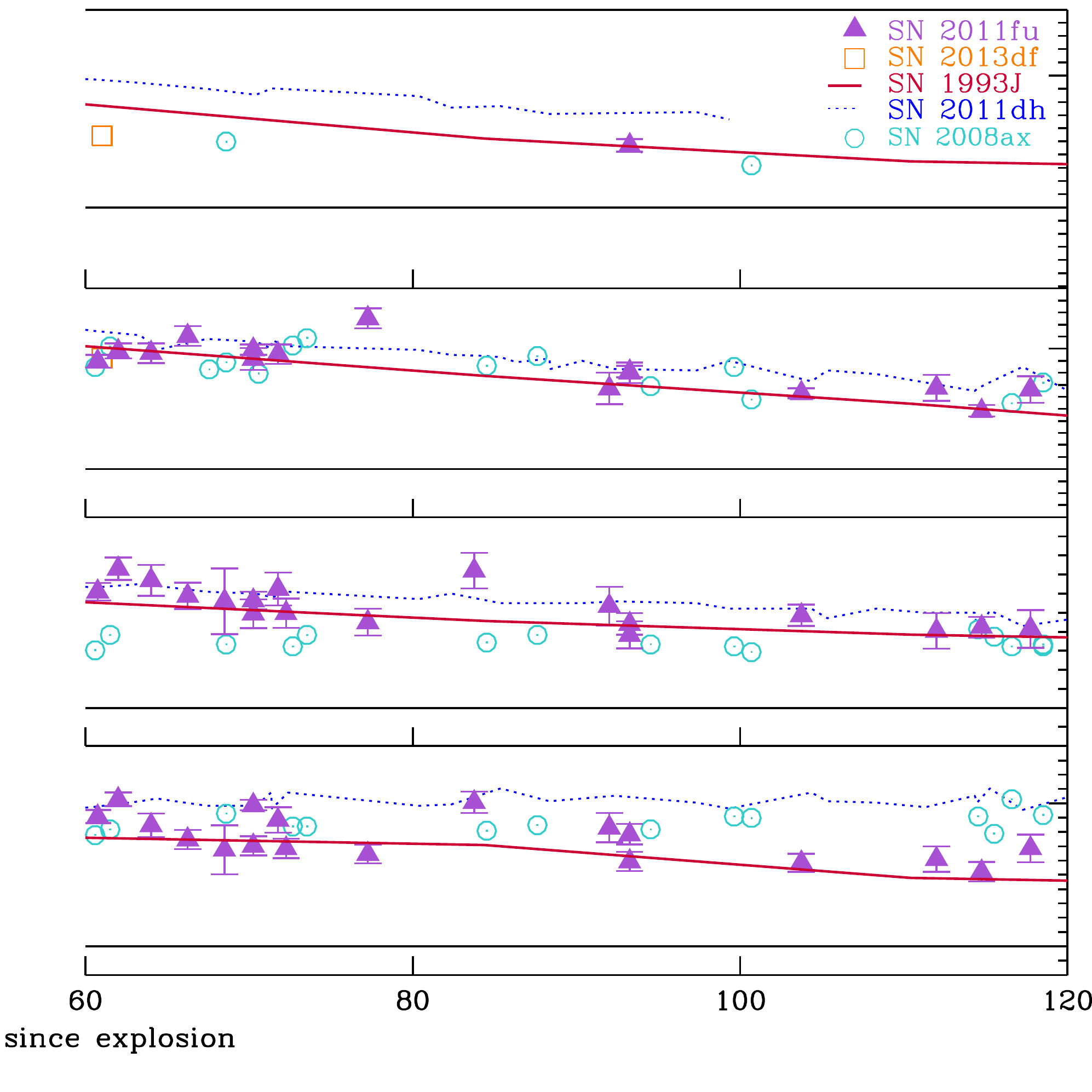}
 \caption{Comparison of the $(U-V)_{0}$, $(B-V)_{0}$, $(V-R)_{0}$, $(V-I)_{0}$ colours of type IIb SNe 1993J, 2008ax, 2011dh, 2011fu and 2013df. The colour of SN~2011fu has been corrected for the assumed reddening $E(B-V)_{{\rm Total}}=0.10^{+0.04}_{-0.03}$  mag, while the data and explosion epochs for the comparison supernovae were taken from the literature. A colour version of this figure can be found in the online journal.}
 \label{fig:colours}
\end{figure*}

Figure \ref{fig:colours} shows the intrinsic colour evolution of SN~2011fu and for comparison the data for other type IIb SNe. such as SN~1993J (\citealt{Lewis94}, \citealt{Barbon95}, \citealt{Richmond94}, \citealt{Matthews2002}; assumed explosion date $\rm JD_{\rm exp}=2449074.0$), SN~2008ax (\citealt{pasto08ax}, \citealt{tau08ax}, \citealt{Tsvetkov09}; assumed explosion date $\rm JD_{\rm exp}=2454528.8$), SN~2011dh (\citealt{Ergon11dhmore100D}; assumed explosion date $\rm JD_{\rm exp}=2455713.0$), SN~2013df (\citealt{13dfMG}; assumed explosion date $\rm JD_{\rm exp}=2456450.0$). The adopted extinctions along the line of site of the SNe plus their distance are the same as those given in Table 7 of \cite{13dfMG}. All colour indices of SN~2011fu 
show a smooth rise from explosion up to $\sim 40$ days, that is $\sim15$~d after secondary peak, and afterwards they have a blue-ward trend.  
\subsection{Pseudo-bolometric LC}
\label{pseudobolLC}

To obtain the pseudo bolometric optical-NIR LC, first we converted the apparent magnitudes of SN~2011fu (corrected for extinction) to effective fluxes. At the phases at which there were no data, the missing points were obtained by interpolation, or as in the cases of the \textit{U} and \textit{u} bands at more than $\sim 90$~d, and NIR prior and past secondary maximum, by extrapolation assuming a constant colour from the data at the nearest epoch. The fluxes were integrated over wavelength following a trapezoidal rule, and finally converted to luminosities taking into account the adopted distance to the SN.

\begin{figure*}
 \includegraphics[width=14cm]{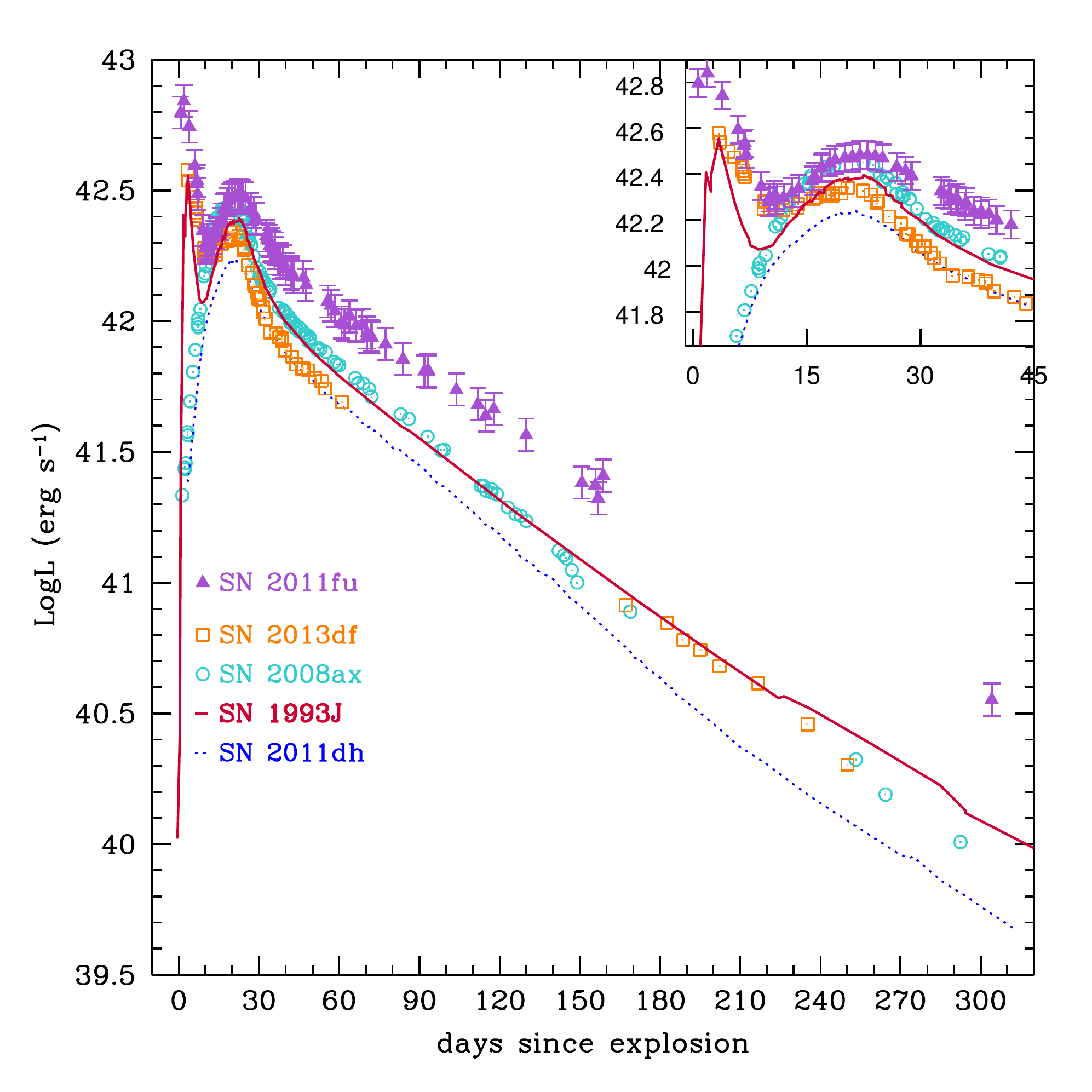}
 \caption{Pseudo bolometric optical-NIR LC of SN~2011fu compared to type IIb SNe 2013df, 1993J, 2008ax, and 2011dh. The data and explosion epochs for the comparison supernovae were taken from the literature. A colour version of this figure can be found in the online journal.}
 \label{fig:bol_other}
\end{figure*}

In Figure \ref{fig:bol_other} we present the pseudo-bolometric NIR-optical LC of SN~2011fu as well as those of SNe 1993J, 2008ax, 2011dh, and 2013df, which we calculated in a similar way. In comparison to SNe~1993J and 2013df, SN~2011fu presents a longer cooling phase after the first peak, and a longer rise time to secondary peak than the rest of the objects, which makes SN~2011fu the one with the highest $^{56}$Ni mass synthesised in the explosion, in accordance with \cite{11fu} and our modelling of the LC, which we present in Section \ref{LCmodelling}. The slope of the LC tail starting at 40~d  and up to $\sim300$~d is a factor 1.1 steeper for SN~1993J, indicating as well that SN~2011fu ejected more mass in its explosion.

\section{Spectroscopy}
\label{spec}
\subsection{Spectral evolution of SN~2011fu}
\label{specevol}
\begin{figure*}
  \includegraphics[width=20cm]{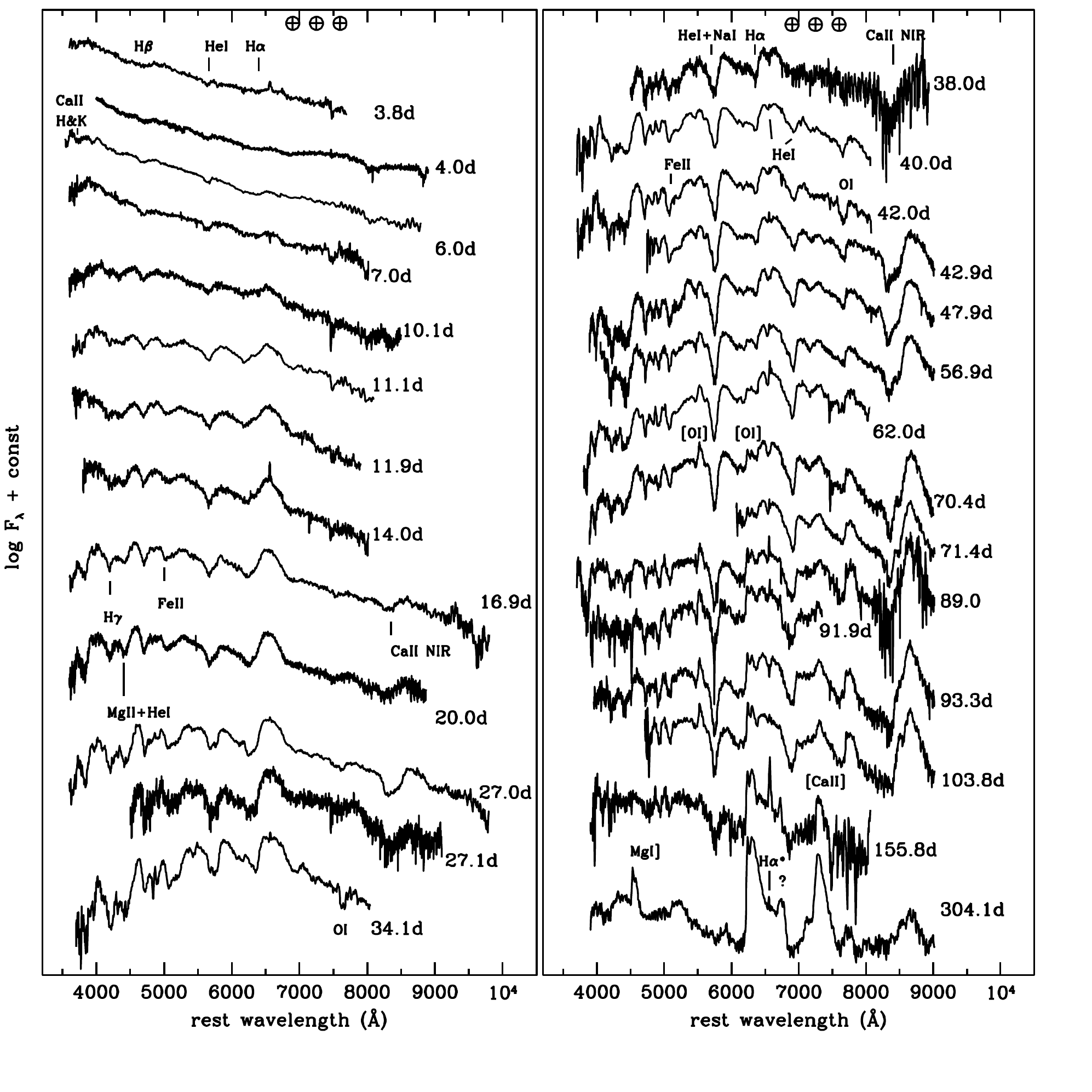}
  \caption{Optical spectral evolution of SN~2011fu, where the most relevant features in the spectra are indicated. The wavelengths at which there are residual telluric features have been  marked with $\oplus$. The spectra have been corrected for the host galaxy redshift. Epochs indicated in the plot are with respect to our assumed explosion date of JD = $2455824.5$ $\pm$ $0.7$. Spectra have been shifted vertically for clarity. H$\alpha$* is the narrow H$\alpha$ emission which we believe is associated to the H \,{\sc ii} region. The question mark at $\sim6700$~\AA{} marks the line we discussed in Section \ref{specevol}.}
  \label{fig:spec}
 \end{figure*}

In Figure \ref{fig:spec} we show the spectral evolution of SN~2011fu ranging from $\sim4$~d to $\sim304$~d after explosion.
The early spectra show a blue continuum and shallow features. Starting at around 6~d  H$\alpha$, H$\beta$ (with some possible Fe\,{\sc ii} contamination), He\,{\sc i} $\lambda$$5876$, Ca\,{\sc ii} H \& K, Ca\,{\sc ii} NIR $\lambda$$\lambda$$\lambda$$8498$, $8542$, $8662$, and Fe\,{\sc ii} $\lambda$$5169$ appear to grow. The H$\alpha$ P-Cygni absorption component at 11.1~d shows a small trough at around 6300~\AA{} in addition to a broader and deeper one centred at approximately 6190~\AA{}. These double H$\alpha$ absorption features have been seen before in some type IIb SNe (e.g. SN~2011hs; \citealt{bufano14}) and have been claimed to be either due to  Si\,{\sc ii} \citep{Hach12} or the presence of a double density distribution of hydrogen in the ejecta of the SNe as claimed for some type II SNe (e.g. \citealt{Inserra13}). Note 
that in the first spectrum  and the one at 14~d, there is a narrow emission line at approximately the rest wavelength of H$\alpha$. This line is also seen in some of our subsequent spectra, however, as we will explain below, we believe it is not associated with the SN or its circumstellar medium (CSM).
After the LC's secondary maximum, at 34.1~d, two new lines with fairly narrow absorptions at approximately 4840~\AA{} and 4890~\AA{} are discernible,  which we believe are associated with Fe\,{\sc ii} $\lambda$4924 and  Fe\,{\sc ii} $\lambda$5018 + He\,{\sc i} $\lambda$5015 respectively. In addition there is an absorption that could be associated with O\,{\sc i} $\lambda$7774.
A major change occurs at $\sim40$~d, when  He\,{\sc i} $\lambda$6678 and He\,{\sc i} $\lambda$ 7065 appear and together with He\,{\sc i} $\lambda$$5876$ become progressively stronger.
 At 62~d a double-peaked emission associated with [O\,{\sc i}] $\lambda$$\lambda$ $6300$, $6364$ with components at approximately 6240 and 6300~\AA{}  is observed, as well as an emission which has 
increased in intensity 
with respect to previous spectra, and is possibly associated with [O\,{\sc i}] $\lambda$5577. Starting at 103.8~d, [N\,{\sc ii}] $\lambda\lambda$6548, 6583 produces an emission between $\sim6400$ and $\sim6700$~\AA{} \citep{lateIIb}. At phase 155.8~d  we notice a decrease in the intensity of Na\,{\sc i} around 5890~\AA{} (which is possibly contaminated by residual He\,{\sc i} $\lambda$5876; \citealt{lateIIb}). In addition we identify [Ca\,{\sc ii}] $\lambda\lambda$$7291$, $7324$ and two strong emission lines, a narrow one at approximately the rest position of H$\alpha$ (on top of N \,{\sc ii} $\lambda$$\lambda$6572, 6583) and the other at $\sim6723$~\AA{}.  In the last spectrum of our sequence, other than [O\,{\sc i}] $\lambda$$\lambda$ $6300$, $6364$, [Ca\,{\sc 
ii}] $\lambda\lambda$$7291$, $7324$, O\,{\sc i}~$\lambda$$7774 $, we also distinguish Mg\,{\sc i}] $\lambda$4571, [Fe\,{\sc ii}]~$\lambda$7155. The Ca\,{\sc ii} NIR line has diminished significantly from our previous spectra and is now blended with [C\,{\sc i}]~$\lambda$8727. The narrow emissions at H$\alpha$ and the one at $\sim6700$~\AA{} are still present although the first seems to have diminished in intensity while the second has become broader and now is centred at approximately 6735~\AA{}. The unresolved narrow H$\alpha$ emission line seen in our two last spectra, and also observed in some of the earlier-time spectra, is detected mainly on nights in which the seeing was not good or the SN was faint, so it is probably due to contamination from a nearby H\,{\sc ii} region.

Concerning the line at $\sim6700$~\AA{}, we note that in the nebular spectra of SN~2013df \citep{13dfMG} there was also an emission line at a similar wavelength, and it was interpreted as a red-shifted component of the SN H$\alpha$ feature due to asymmetrical ejecta-circumstellar material interaction. However, since the line is detected in two different SNe it may be linked to a specific transition. In the models presented by \cite{lateIIb} there is an emission associated with [S\,{\sc ii}] $\lambda\lambda$6716, 6731 about an order of magnitude weaker than the observed line. However, uncertainties in the temperature and ionization state might cause it to be stronger for SN~2011fu.  
  Another possibility is due to contamination by the nearby H\,{\sc ii} region which we believe is causing the narrow H$\alpha$ emission.  In fact, in the 156~d spectrum both the narrow H$\alpha$ and the $\sim6700$~\AA{} feature have similar FWHM  and are unresolved suggesting a common origin not related to the SN. Similar narrow H$\alpha$ and [S\,{\sc ii}] $\lambda\lambda$6716, 6731 emissions caused by nearby H~\,{\sc ii} regions are detected e.g. in the spectra of the type IIb SNe~1987K \citep{Fpk88} and 2008ax \citep{Milisavljevic10}. However, in H\,{\sc ii} regions, [S\,{\sc ii}] $\lambda\lambda$6716, 6731 lines are less intense than H$\alpha$ and this is not the case of the 304~d spectrum of SN~2011fu. In addition, the line at $\sim6700$~\AA{} at this phase is clearly resolved, for these reasons, a [
S\,{\sc ii}] $\lambda\lambda$6716, 6731 emission associated with the SN is the favoured possibility at 304~d. \\

\begin{figure}
 \includegraphics[width=8.6cm]{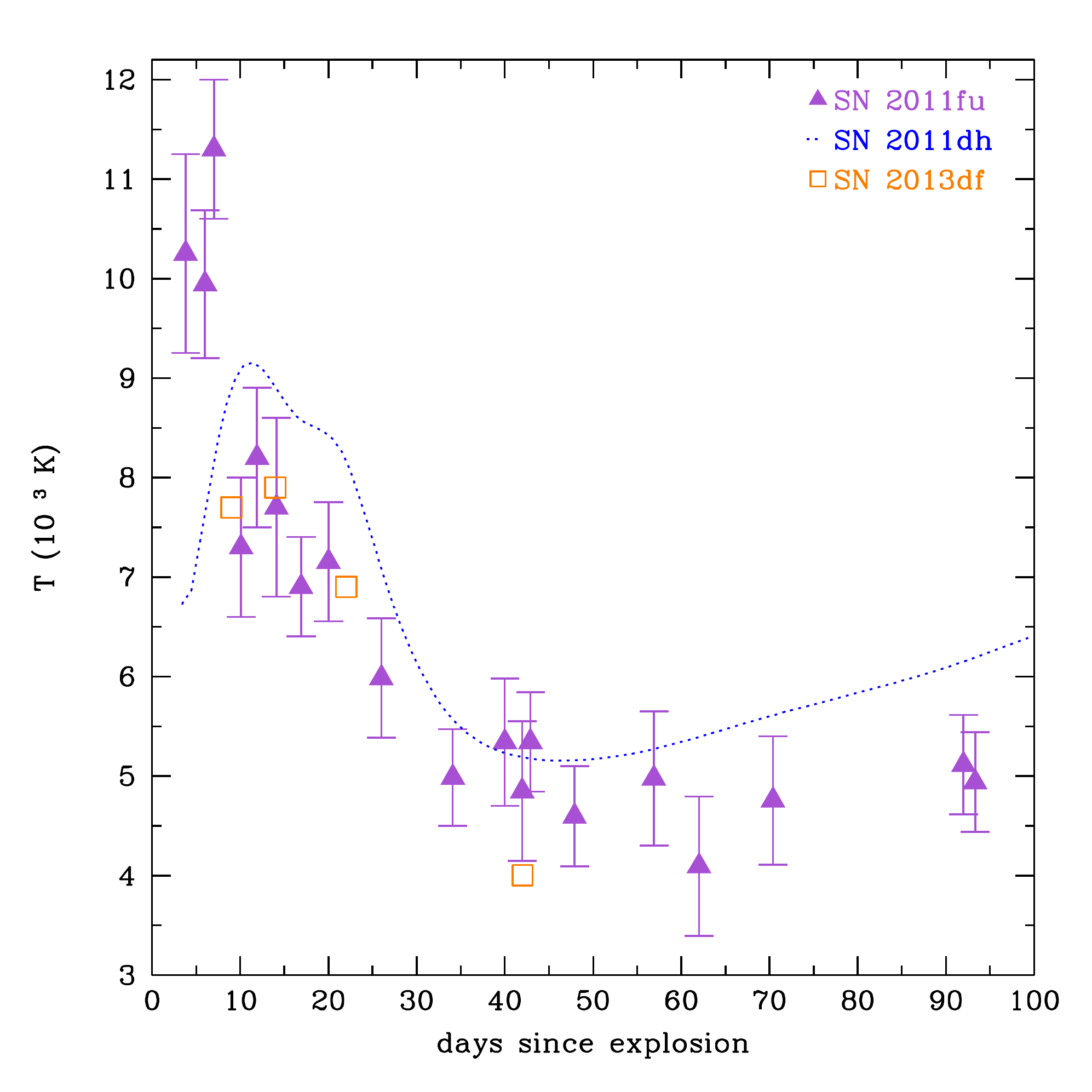}
 \caption{Black body temperature evolution of SN~2011fu with respect to our assumed explosion date, JD = $2455824.5$ $\pm$ $0.7$, compared to those of SNe~2011dh and 2013df. A colour version of this article can be found in the online journal.}
 \label{fig:blackbody}
\end{figure}

We have estimated the black-body temperatures of the SN by fitting a black-body function to the whole spectrum at each epoch. As can be seen in Figure \ref{fig:blackbody}, the black-body temperature evolves with a steep decline up to approximately 40 days since explosion and after that, it roughly remains constant.  The time evolution of the temperatures is in agreement with the colour evolution of the SN and with that derived by \cite{11fu}. As seen in Section \ref{colours}, the colours redden during approximately the first 40 days and after that they are practically constant with a slight trend towards the blue. In the first $\sim10$~d, the trend of the black body temperature is opposite to SN~2011dh \citep{Ergon11dh100D}. Note that during this phase the LC of SN~2011fu is in the adiabatically-powered declining phase to the minimum after the first peak while the pseudo bolometric LC of SN~2011dh is radioactively-powered.

\subsection{Velocities}

\begin{figure*}
 \hspace{-0.5cm}
 \includegraphics[width=6.2cm]{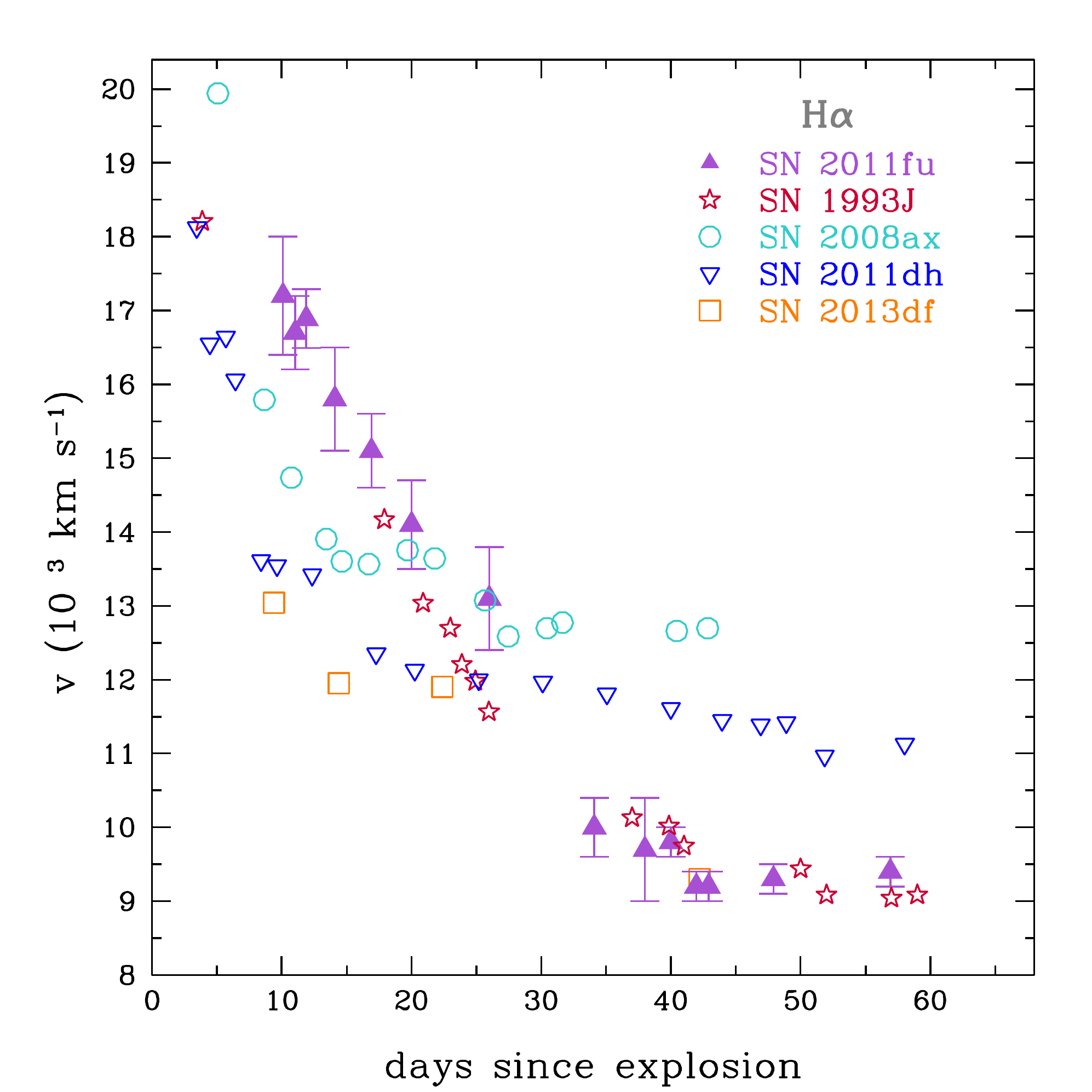}
 \hspace{-0.5cm}
 \includegraphics[width=6.2cm]{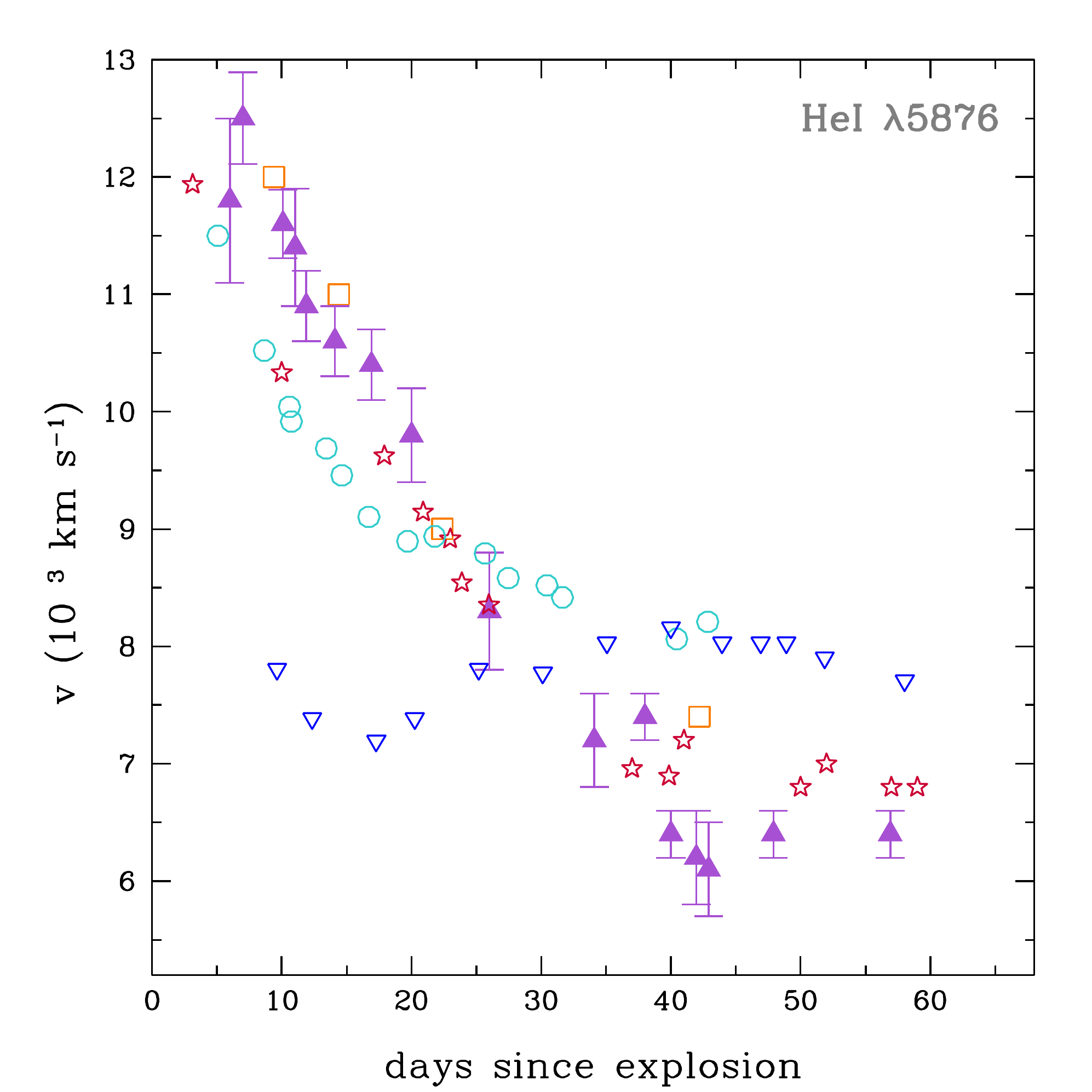}
 \hspace{-0.5cm}
 \includegraphics[width=6.2cm]{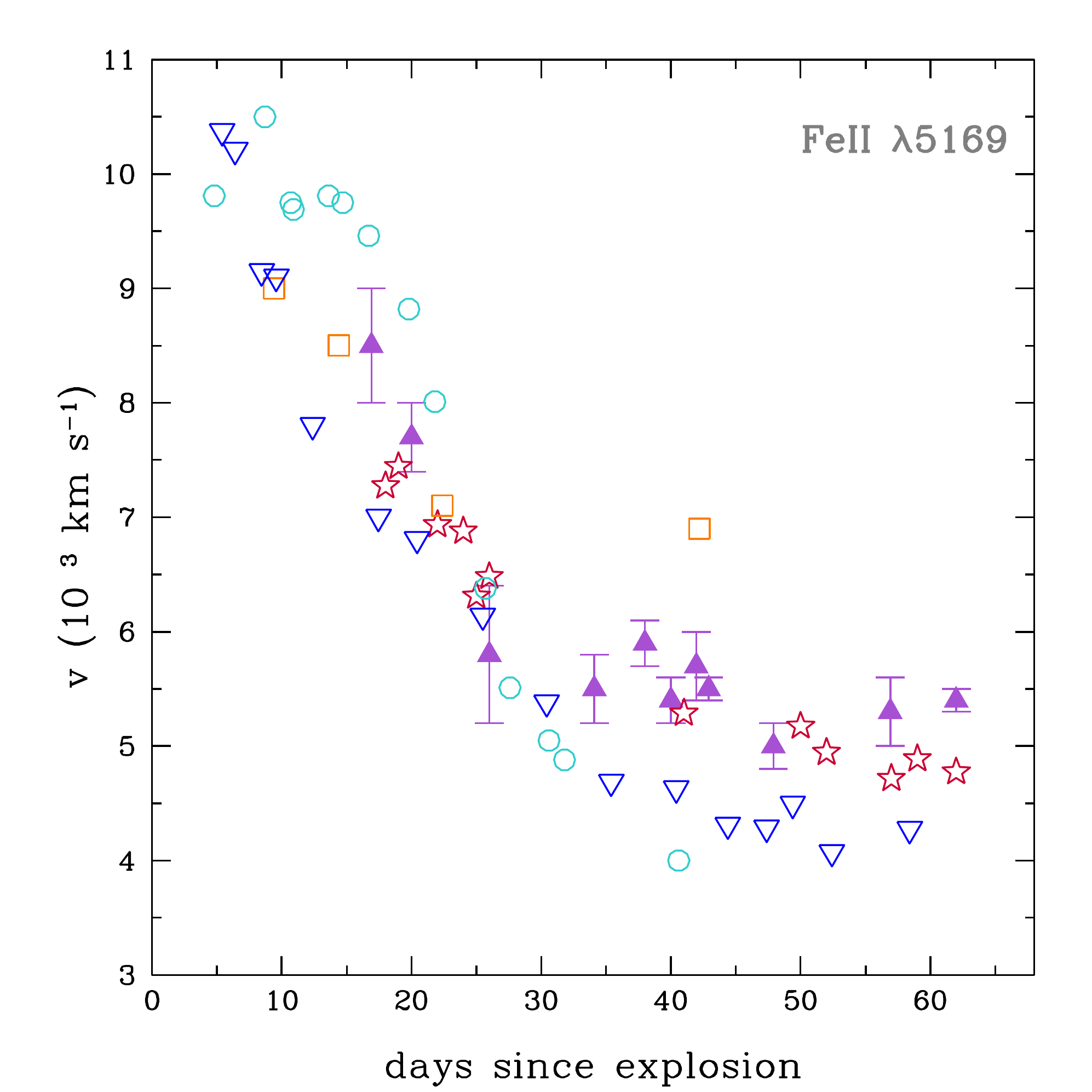}
 \caption{Velocity evolution of H$\alpha$, He\,{\sc i} $\lambda5876$, and Fe\,{\sc ii} $\lambda$$5169$ for SN~2011fu compared to those of SNe 1993J, 2008ax, 2011dh, and 2013df.}
 \label{fig:velocities}
\end{figure*}

From the minima of the P-Cygni absorptions of H$\alpha$ and He\,{\sc i} $\lambda5876$, we have an estimate of the velocities of the ejected material. Our results are presented in Figure~\ref{fig:velocities} together with the velocities for SNe~1993J \citep{Barbon95,tau08ax}, 2008ax \citep{tau08ax}, 2011dh \citep{Ergon11dh100D}, and 2013df \citep{13dfMG}. Note that the first two He\,{\sc i} $\lambda5876$ velocities reported in the figure were obtained from shallow, low contrast profiles, and in consequence are uncertain. 

As we can see in  Figure \ref{fig:velocities}, both the H$\alpha$ and He\,{\sc i} $\lambda5876$ velocities we have derived are overall similar to those for the comparison SNe. Between 12~d and 28~d we note that the H$\alpha$ velocities for SN~2011fu are around 2000~km~s$^{-1}$ higher than those obtained with SYNOW by \cite{11fu}. From $\sim 40$~d on, the velocities remain constant and similar to those of SN~1993J but lower than the ones derived for SN~2011dh. We would like to point out that around a month past explosion the He\,{\sc i} $\lambda5876$ absorption component has a complex profile with a double trough, however, also in these cases the values reported in the figure are those obtained by adjusting a single Gaussian to the whole profile. Thus, the velocities measured for He\,{\sc i} $\lambda5876$, may have an added uncertainty due to the possible contamination by Na~\,{\sc i}
~$\lambda\lambda5890$,~$5896$, especially at later times. Noteworthy is the fact that for SN~2011dh, the hydrogen lines were never seen below $11000$~km~s$^{-1}$, whereas the helium lines were always constrained to lower velocities. In fact the modelling of the SN data suggested that $11000$~km~s$^{-1}$ marked the interface between the H-rich envelope and the He core. For SN~2011fu, however, we cannot establish an analogous boundary based solely on the observed velocities, since in the velocity space H$\alpha$ is not separated from He.

The photospheric velocities are expected to be similar to those derived from the Fe\,{\sc ii} $\lambda$$5169$ line (see e.g. figure 14 of \citealt{DessartHillier2005}), these are shown in the right panel of Figure \ref{fig:velocities}. As can be seen in this figure the velocities of Fe\,{\sc ii} for SN~2011fu are overall comparable with those of our selected sample of type IIb SNe. We notice again that up to 30~d they are lower than the ones derived by \cite{11fu} by about 2000~km~s$^{-1}$. 

In summary, the expansion velocities of SN~2011fu are similar to those of other type IIb SNe.

 \begin{figure*}
   \includegraphics[width=15.5cm]{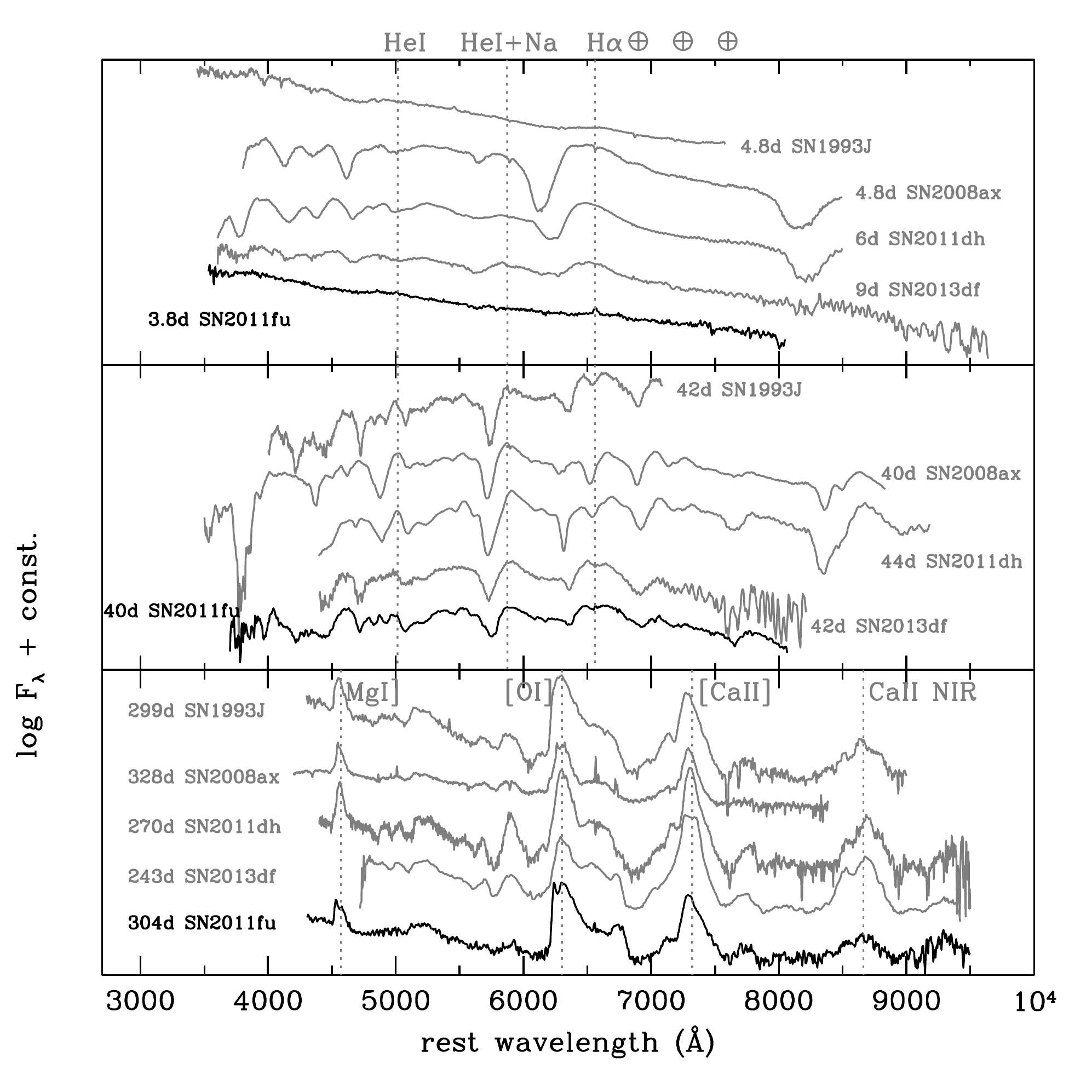}
 \caption{Comparison of early (5-9 days since explosion), intermediate (40-44 days after explosion), and late (299-398 days after explosion) spectra of SN~2011fu with those of type IIb SNe 1993J, 2008ax, 2011dh, and 2013df. The original references for the spectra of the comparison SNe are: \citealt{Barbon95} for SN~1993J at 4.8 and 42~d, and an unpublished spectrum taken at the 1.82-m telescope in Asiago (Italy); \citealt{tau08ax} for SN~2008ax at 4.8 and 40~d, \citealt{Mdj14} for SN~2008ax at 328~d; \citealt{Ergon11dh100D} for SN~2011dh at 6 and 44~d, \citealt{shivvers2013} for SN~2011dh at 270~d, and \citealt{13dfMG} for SN~2013df's spectra. The spectra have been dereddened, redshift corrected and shifted vertically for clarity.}
  \label{fig:spec_comparison}
  \end{figure*}

\subsection{Comparison to other type IIb SNe}
\label{spec_comparison}
In Figure \ref{fig:spec_comparison} we present a comparison of spectra of SN~2011fu at three different phases in its evolution along with those of other type IIb SNe downloaded from WISEREP\footnote{http://wiserep.weizmann.ac.il/spectra/list} \citep{yaron12}. As can be seen in the top panel of the figure,  the spectra of SNe~1993J, 2011fu and 2013df exhibit a blue almost featureless continuum with shallow hydrogen and helium lines (note that the spectrum of SN~2013df corresponds to a later phase). The coeval spectra of SNe 2008ax and 2011dh show much stronger lines and redder continuum, reflecting the absence of the first peak of the LC. In the middle panel of Figure \ref{fig:spec_comparison} we have depicted a spectral comparison at an intermediate phase ($\sim40$~d). Although there is an overall resemblance in the spectra, the intensity of 
the lines varies from one SN to the other, indicating differences in the temperature of their ejecta. Specifically, SN~2011fu has the weakest H$\alpha$ and 
He\,{\sc i} $\lambda$$5876$ absorptions at this phase. Although not shown in the figure, we have also compared a spectrum of SN~2011fu at 22~d with the other type IIb SNe at similar phases. Again, we have found the intensity of the H$\alpha$ and He\,{\sc i} $\lambda$$5876$ lines to be lower than for the comparison SNe.
For the nebular spectra (bottom panel of Figure \ref{fig:spec_comparison}), in the case of SN~2011fu, the [O\,{\sc i}] $\lambda$$\lambda$ $6300$, $6364$ emission profile is double-peaked, as it is for SN~2008ax. The Mg\,{\sc i}] $\lambda$4571 profile for SN~2011fu is also double-peaked and again more similar to that of SN~2008ax. Interestingly and as noted above, SN~2011fu shows an emission line at $\sim 6700$~\AA{}  similarly to SN~2013df, although  the line is broader in SN~2011fu.


\subsection{Line profiles}

\begin{figure*}
   \includegraphics[width=11.5cm]{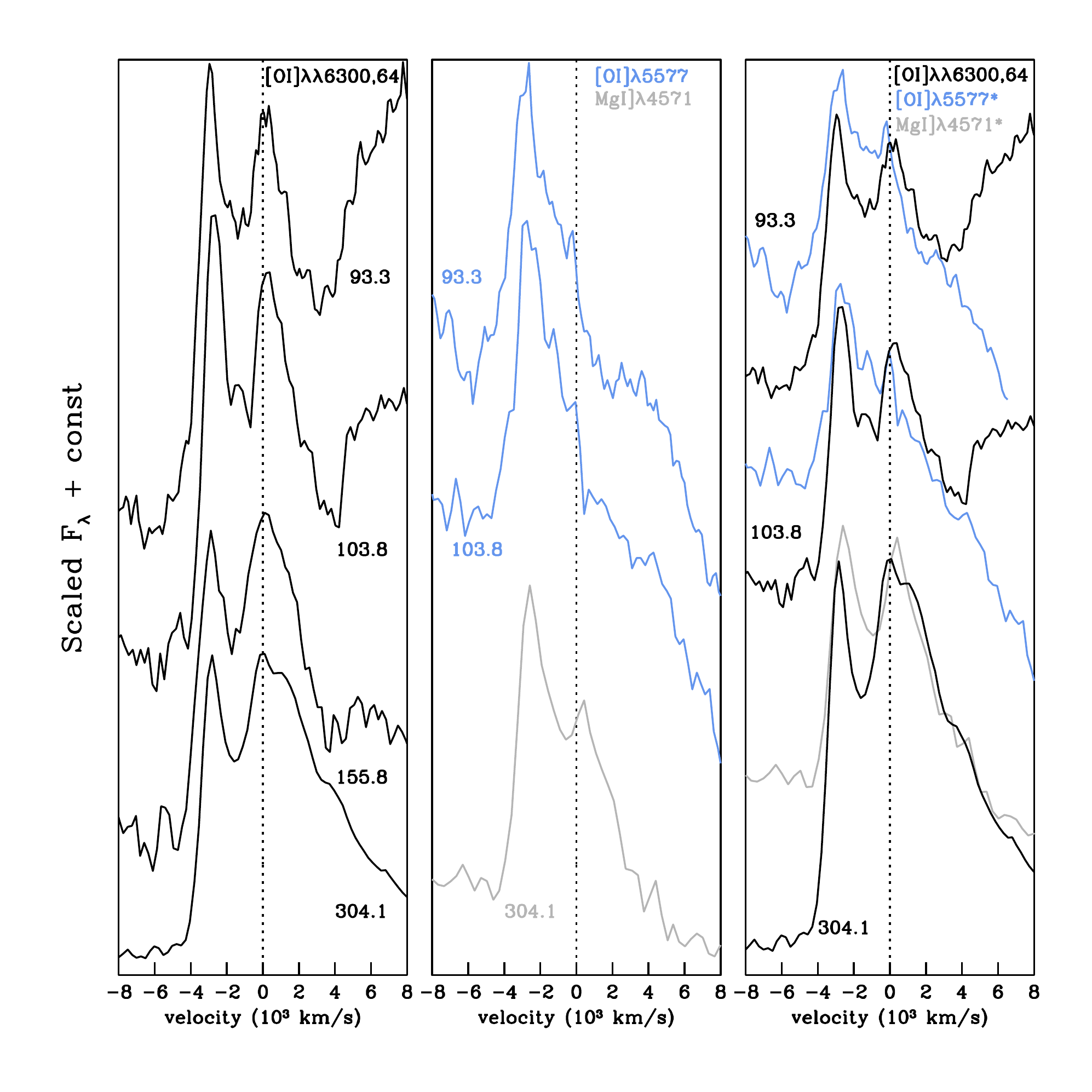}
 \caption{Left panel: evolution of the late-time profiles of [O\,{\sc i}]~$\lambda$$\lambda$$6300$, $6364$ in velocity space. The 0~km~s$^{-1}$ vertical line corresponds to 6300~\AA{}. Middle panel: [O\,{\sc i}]~$\lambda$$5577$ and Mg\,{\sc i}] $\lambda$4571 profiles between 93 and 304~d since explosion. Right panel: Comparison of the [O\,{\sc i}]~$\lambda$$\lambda$$6300$, $6364$ profiles with [O\,{\sc i}]~$\lambda$$5577$* and Mg\,{\sc i}]~$\lambda$4571*, which are the original profiles plus an artificial component, scaled to 1/3 the intensity and redshifted by 3000~km~s$^{-1}$, added to account for the doublet nature of [O\,{\sc i}]~$\lambda$$\lambda$$6300$, $6364$.}
  \label{fig:oxygen_profiles}
  \end{figure*}

In the left panel of Figure \ref{fig:oxygen_profiles} we present the evolution of SN~2011fu's [O\,{\sc i}] $\lambda$$\lambda$ $6300$, $6364$ nebular profile in velocity space. As seen in the figure, the profile shows two peaks, one at approximately 0 km~s$^{-1}$ and the other at $\sim -2800$~km~s$^{-1}$. In the middle panel of Figure \ref{fig:oxygen_profiles} we show the nebular profiles of [O\,{\sc i}] $\lambda$$5577$ and  Mg\,{\sc i}] $\lambda$4571. Finally, in the right panel of the figure we have artificially added a component with 1/3 the intensity of the original [O\,{\sc i}] $\lambda$$5577$ and  Mg\,{\sc i}] $\lambda$4571 profiles and redshifted by 3000~km~s$^{-1}$, in order to compare them with the [O\,{\sc i}] $\lambda$$\lambda$$6300$, $6364$ doublet. As can be seen in this last panel, the shapes of the oxygen and magnesium profiles are quite similar to one another.  \cite{tau09} already found a great similarity between the oxygen and magnesium profiles in nebular spectra of some 
 stripped envelope 
SNe, supporting the idea that Mg and O have similar spatial distributions within their ejecta, which is expected from the models \citep{Maeda06}. Moreover, in the case of the type IIb SN~2011dh the components of the small scale fluctuations of the [O\,{\sc i}] $\lambda$$\lambda$$6300$, $6364$ and Mg\,{\sc i}] $\lambda$4571 in its late-time spectra coincide, which in addition to the results of \cite{lateIIb} imply that the oxygen lines are mostly formed in the O/Ne/Mg zone.

One possibility to explain the shape of the profiles of SN~2011fu is that the bulk of oxygen and magnesium is distributed in spherically symmetric expanding ejecta but that there is a clump with emission from both these elements expanding at  $\sim 2800$~km~s$^{-1}$ towards the observer. A second  possibility is that the radioactivity exciting the lines is distributed asymmetrically. Given the similarity of the [O\,{\sc i}]~$\lambda$$\lambda$$6300$, $6364$ and the [O\,{\sc i}]~$\lambda$$5577$ and  Mg\,{\sc i}]~$\lambda$4571 when an artificial component is added to the last two, an H$\alpha$ high-velocity absorption, as found by \cite{Maurer10} for some type IIb SNe, is not likely the principal cause of the double-peaked [O\,{\sc i}]~$\lambda$$\lambda$$6300$, $6364$ line, although it could be contributing to the profile. On the other hand, the narrow blue-shifted component of the [O\,{\sc i}]~$\lambda$$\lambda$$6300$ carries a significant fraction of the flux even at 300~d. Residual opacity in the core or 
line 
blocking 
explains blue shifted emission lines of stripped envelope  and  specifically type IIb SNe (\citealt{tau09}; \citealt{lateIIb}). But in the case of SN~2011fu the persistence of the blue shifted emission from the [O\,{\sc i}]~$\lambda$$\lambda$$6300$, $6364$ line over time suggests some degree of asymmetry in the SN ejecta. So, in principle, either clumping or an asymmetrical distribution of the radioactive material seem the most likely explanations for the profiles.
Intriguingly, we have found two other SNe in the literature that have blue shifted double-peaked [O\,{\sc i}]~$\lambda$$\lambda$$6300$, $6364$ profiles similar to SN~2011fu: Types Ib SN~1996aq \citep{tau09} and SN~2005bf \citep{Mdj08,Milisavljevic10} at 216d. For SN~1996aq \cite{tau09} speculated that  the most likely explanation for its profile shape is that a clump is moving in the line of site at high velocity, and this is probably also the case for SNe~2011fu and 2005bf.

Finally, we note that the shape of SN~2011fu's [Ca\,{\sc ii}]~$\lambda \lambda$$7291$, $7324$ line (not shown in the figure) is not similar  to the oxygen and 
magnesium profiles, implying that, as expected, they are forming at different locations within 
the ejecta.

\section{Discussion}
\label{discussion}

\subsection{Hydrodynamical modelling of the pseudo-bolometric LC}
\label{LCmodelling}

One approach to derive physical parameters of SNe is to compare
LCs and expansion velocities derived from hydrodynamical models with
observations. From this comparison it is possible to constrain the
explosion energy, the ejecta and the nickel masses as well as the
distribution of the radioactive material. These parameters can be
estimated by focusing the modelling around the ``main peak'' of the LC (i.e.
the nickel powered phase). Information on the
size of the progenitor is given by the post shock breakout cooling
phase. \\

We calculated a set of hydrodynamical models using as initial
structures those derived from stellar evolutionary calculations. A one-dimensional
Lagrangian LTE radiation hydrodynamics code \citep{bersten11} is
used to explode the initial configuration. The code allows to
calculate consistently the shock wave propagation in the stellar
interior, the shock breakout and the posterior phases of the LC evolution
until the SN becomes nebular. An update of the micro physics (equation of
state and opacities) appropriate for the study of stripped-envelope SNe
were incorporated in the code by \cite{Melina12}. As pre-SN
structure we adopted helium (He) stars with a thin hydrogen envelope ($<
1 \, M_\odot$) which have successfully reproduced the LCs and the spectral
features of other SNe IIb (see e.g. \citealt{1994ApJ...420..341S}, \citealt{1998ApJ...496..454B}, and more recently \citealt{Melina12}). The He core
models used here were calculated by \cite{NH88} following
the stellar evolution until the collapse of the core.  We have
smoothly attached a low-mass hydrogen-rich envelope in hydrostatic and
thermal equilibrium to the He core to take into account the thin
H-envelope required for a SN IIb. \\

\begin{figure*}

   \includegraphics[width=6.5cm, angle=270]{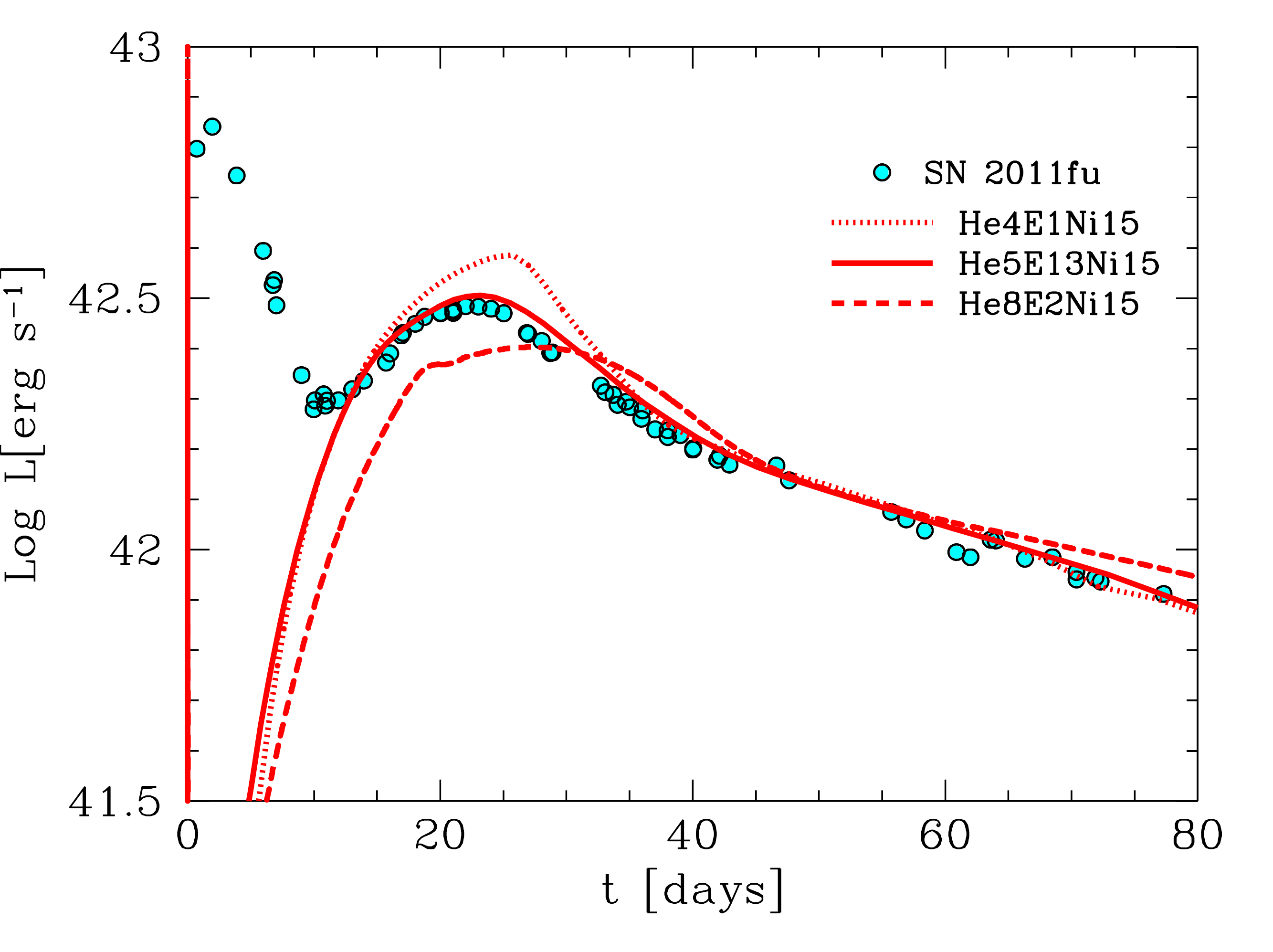}
   \includegraphics[width=6.5cm, angle=270]{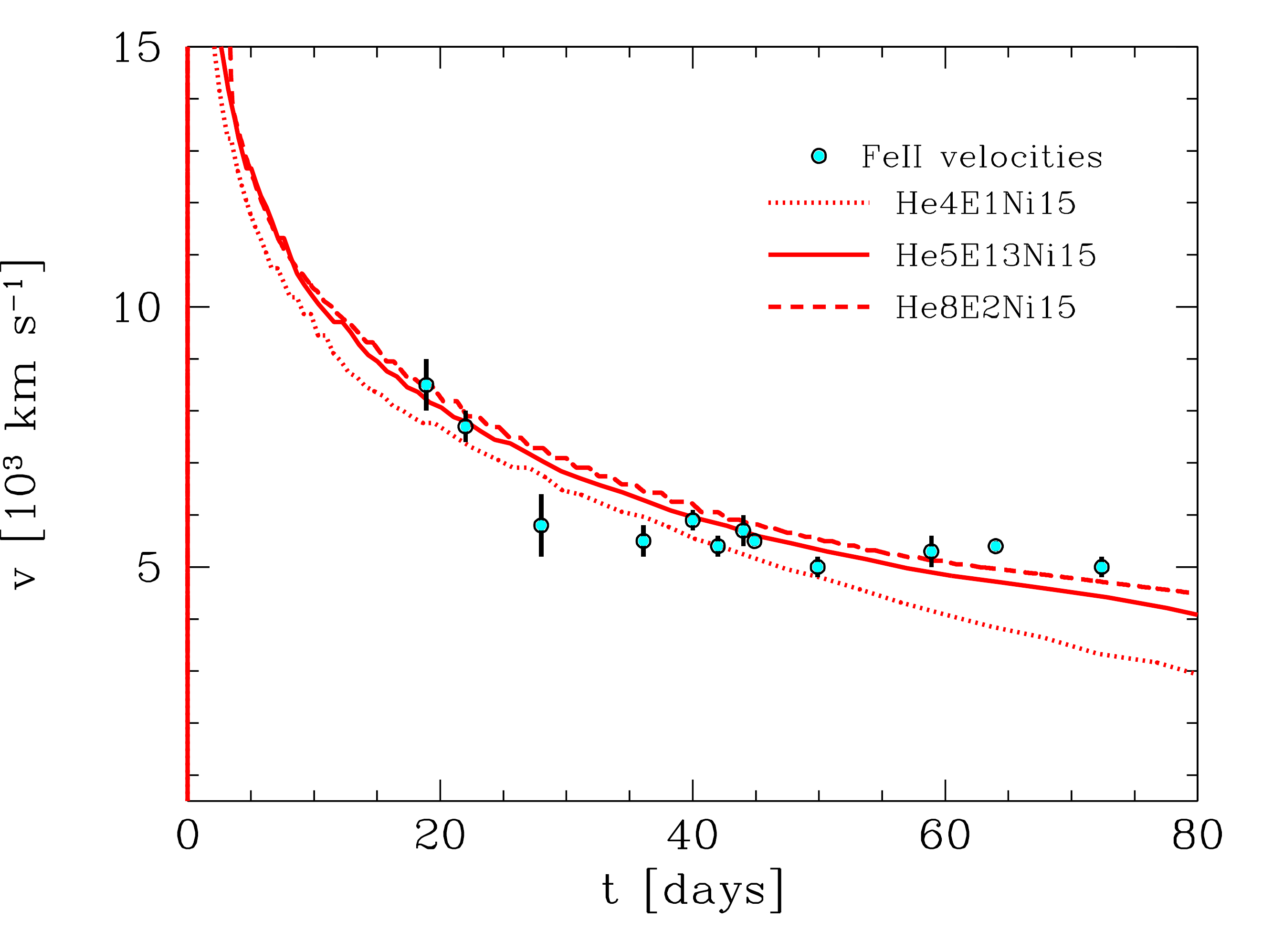}
 \caption{Left panel: Observed bolometric light
  curve of SN~2011fu (dots) compared with the results of the light
  curve calculations for models He4E1Ni15 (dotted line), He5E13Ni15 
(solid line) and He8E2Ni15 (dashed line) omitting the first peak. Right panel:
  Evolution of the photospheric velocity for models He4E1Ni15 (dotted
  line), He5E13Ni15  (solid line) and He8E2Ni15 (dashed line) compared
  with measured Fe~II line velocities of SN~2011fu (dots).}
  \label{11fu_model_mass}
  \end{figure*}

First we focus our analysis on the modelling of the LC and the
photospheric velocities around the secondary peak without taking into
account the cooling part of the LC. Figure~\ref{11fu_model_mass} shows the results of the LC modelling for three different He core masses,
4 $M_\odot$ (He4), 5 $M_\odot$ (He5), and 8 $M_\odot$ (He8), which
correspond to the stellar evolution of single stars with main-sequence
masses of 15, 18, and  25 $M_\odot$, respectively. For each 
model different explosion energies and $^{56}$Ni masses were explored. Here
we show the best set of parameters for each of the
models. Specifically, an explosion energy of 1, 1.3 and 2 foe (1~foe~=~
1 $\times 10^{51}$ erg) and a $^{56}$Ni mass of 0.15 M$_\odot$\footnote{We have estimated the uncertainty of the $^{56}$Ni mass by considering  only the photometric errors to be  0.02~M$_\odot$} were
found for model He4 (He4E1Ni15), He5 (He5E13Ni15) and He8 (He8E2Ni15) respectively. 
As can be seen in the figure, model He5 provides the best
representation of the observed data. The He8 model is too massive to reproduce the LC
unless we assume a more energetic explosion but this would 
fail to fit the expansion velocities. On the other hand a
lower mass model, He4, gives a worse fit of the LC and  underestimates
the early photospheric velocities.  \\

\begin{figure}

   \includegraphics[width=6.5cm, angle=270]{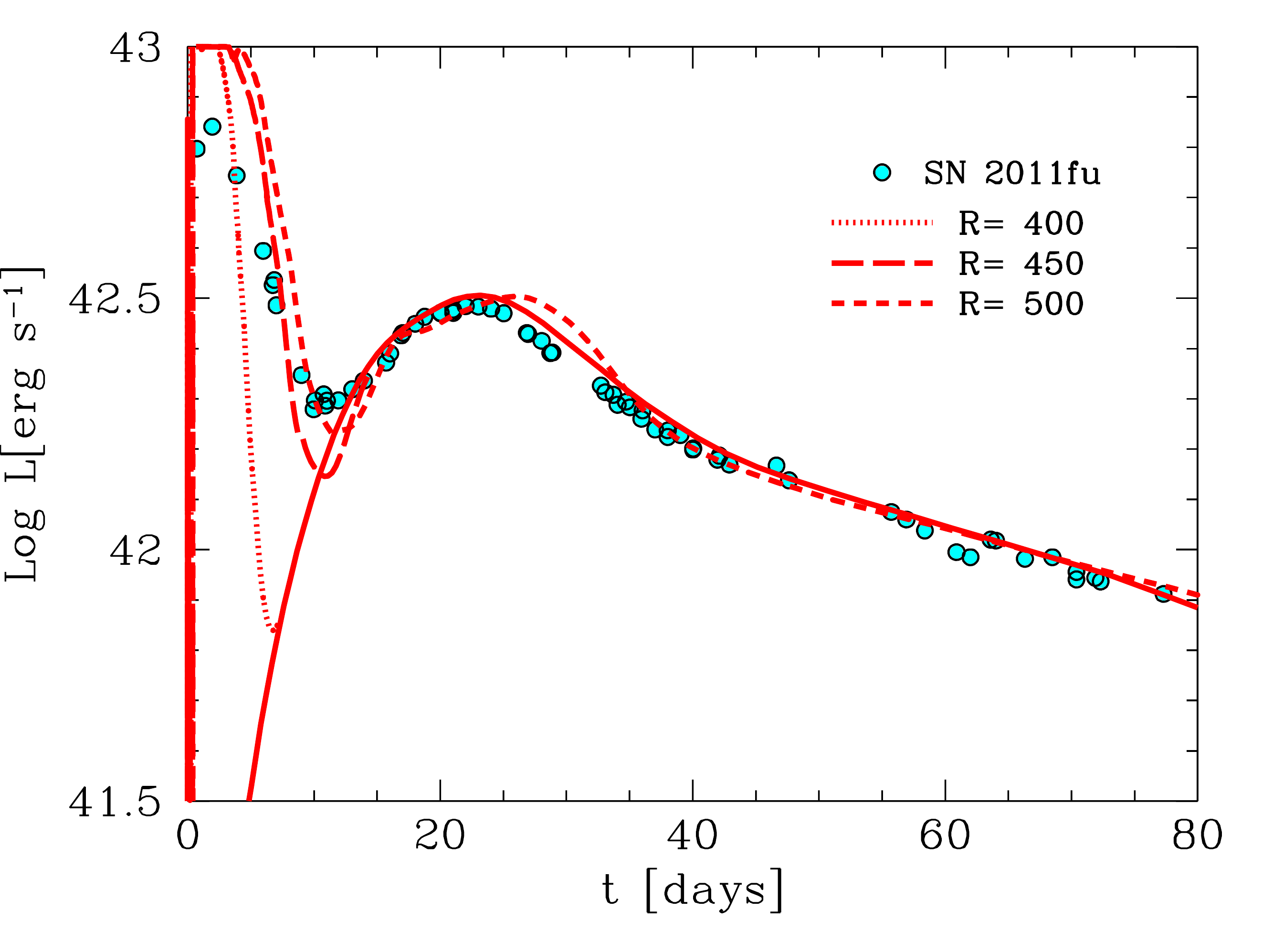}
 \caption{Observed bolometric LC of SN~2011fu
    (dots) compared with the bolometric LCs for models with the same
    physical parameters than our preferred model (He5E13Ni15; solid line), but
    different initial radii. The radius variation is
    accomplished by attaching thin H-rich envelopes to the He5 model. 
    An extended progenitor with $R \approx 450 \,R_\odot$ (long dashed) is needed to
    reproduce the cooling phase of SN~2011fu.}
  \label{11fu_envelope}
  \end{figure}

Our favoured model, He5, is a H-free object with an initial
radius of $\approx$ 2 R$_\odot$. Although, this model gives a good
representation of the data around the secondary peak, it fails to reproduce 
the cooling phase due to the compact structure of the
progenitor. Figure~\ref{11fu_envelope} shows the fits to the LC for models
with different progenitor radius. The size of the progenitor was
modified by attaching thin H-rich envelopes to the core of the He5 model (solid line). 
Note that the presence of a thin envelope mostly affects the shape of
the LC during the cooling phase. Figure~\ref{11fu_envelope}
shows models with radii of 400, 450, 
and 500~R$_\odot$ and envelope
masses ($M_\mathrm{env}$) of $\approx$ 0.3 M$_\odot$. The model with
$R=450$ $R_\odot$ provides the closest match to the  
data. However, all the models give a initial
peak brighter than the observations. The differences may be due to uncertainties in the bolometric
calculation (note that we have no UV data for the SN, which are important at these early phases) or/and limitations of the model such as the LTE
approximation or variation of the density profile of the outermost layers, among other. 
It is noteworthy that to reproduce the early phase of the LC
it was not only needed to assume an extended envelope but  
it was also  necessary  to adopt a slightly more massive envelope than in
previous modelling of SNe~IIb, where e.g. $M_\mathrm{env}= 0.1$~$ M_\odot$
was required for SN~2011dh and  SN~1993J (\citealt{Melina12}; see also \citealt{NkPiro14}).
The need for a more massive envelope was mainly due to the high
luminosity of the minimum after primary maximum of the LC. For lower $M_\mathrm{env}$ the cooling
occurs faster and the minimum occurs at a lower luminosity even for
large radius.   \\ 
A summary of all the parameters obtained from the  best core and envelope model that fit the observed LC is shown in Table~\ref{explosion_param}. Similar to  \cite{11fu} we have obtained a kinetic energy which is relatively higher than that obtained for other type IIb SNe but still lower than those derived for SN~2011ei \citep{Mili12} and the type IIb hypernova SN~2003bg \citep{Hamuy09}. The comparison to the pseudo-bolometric LCs of other type IIb SNe in Section \ref{pseudobolLC} indicates that SN~2011fu synthesised more $^{56}$Ni in its explosion than SNe 1993J, 2008ax, 2011dh and 2013df (cf table 7 of \citealt{13dfMG}). The modelling of the LC corroborates this result. Note that we obtain a lower $^{56}$Ni mass than \cite{11fu} (0.21~M$_{\odot}$) possibly due to the different distance and extinction to SN~2011fu that we have adopted while our estimate of the ejected mass is larger than theirs (1.1~M$_{\odot}$).\\
 
 Concerning the progenitor radius, core mass, and hydrogen envelope, we have obtained overall larger values than those obtained by \cite{11fu} with analytical models. Specifically they obtained a progenitor radius of $\sim150$~R$_{\odot}$, a He core mass of $1$~M$_{\odot}$, and a hydrogen envelope mass of $0.1$~M$_{\odot}$. But as they noted, their results should be considered only order of magnitude estimates. With hydrodynamical modelling, we obtained a radius that is 3 times larger and consistent with that of an extended supergiant similarly to the progenitors of SNe~1993J ($\sim 600$~R$_{\odot}$, \citealt{Mnd04,VD13}) and 2013df ($\sim 550$~R$_{\odot}$, \citealt{VD13c}). All in all, our calculations show that the progenitor of SN~2011fu was not a Wolf Rayet (WR) star but a supergiant.

\begin{table*}

\centering
\caption{Explosion parameters and progenitor properties for the best fitting models to the observed data of SN~2011fu.}
\begin{tabular}{cc}
\hline
Parameters & \\
 \hline 
  $E_{\rm kin}$ ($10^{51}$erg)      & 1.3 \\
  $^{56}$Ni mass (M$_{\odot}$)   &   $0.15$      \\
 $M_{\rm ej}$ (M$_{\odot}$)  &    3.5*           \\
 Progenitor Radius (R$_{\odot}$) & 450        \\
Initial mass (M$_{\odot}$) &   18           \\
Hydrogen envelope mass  (M$_{\odot}$)& 0.3\\ 
  \hline

\end{tabular}

\begin{flushleft}
 *Assuming that 1.5 M$_\odot$ collapsed to form a compact remnant and the rest of the mass that formed the core was ejected.
\end{flushleft}

\label{explosion_param}
\end{table*}

\subsection{Comparison with late time spectral models}
\label{spectralmodels}

With the objective of better understanding the nature of the progenitor of SN~2011fu and its ejecta, we have compared the last three spectra of our sequence (104, 156, 304~d)  with the late time spectral models presented in \cite{lateIIb}. 

To begin with we compared our spectra with the three models with the same (best-fit
to SN~2011dh)
settings for mixing, clumping, molecule, and dust, and differing only in progenitor
mass (12, 13 and 17 M$_{\odot}$).
These are models 12C, 13G, and 17A (see table 4 in \citealt{lateIIb}). In order to do this, we scaled the models accounting for differences in $^{56}$Ni mass synthesised in the explosion, phase (if the difference in phase between models and spectra is $\Delta t$, the factor $\rm exp (-2\Delta t/111.4)$ is used to scale the models), and distances assumed for the models and the SN. The models have been calculated at a distance of 7.8 Mpc and assuming a $^{56}$Ni mass of  0.075~M$_{\odot}$ (since they were constructed to compare with SN~2011dh), while in Section \ref{LCmodelling} we have estimated  $0.15\pm0.02$~M$_{\odot}$ of $^{56}$Ni from SN~2011fu.  Overall the model spectra are dimmer than the SN spectra. We note that for $^{56}$Ni on the higher end of its uncertainty the flux difference between modelled and observed 
spectra diminishes. Given that the errors in the extinction are not large, it is likely not the cause of the discrepancy between the flux levels. 
Another 
possibility is that the ejecta structure of SN~2011fu is 
different than the one used in the models and the IIb SNe fit with the models. Note that the  \cite{lateIIb} models have an imposed dust extinction of $\tau$=0.25 from
200~d,
if no dust is produced in SN~2011fu this introduces a 25$\%$ flux error for the
last spectrum.  

\begin{figure*}
 \includegraphics[width=15.5cm]{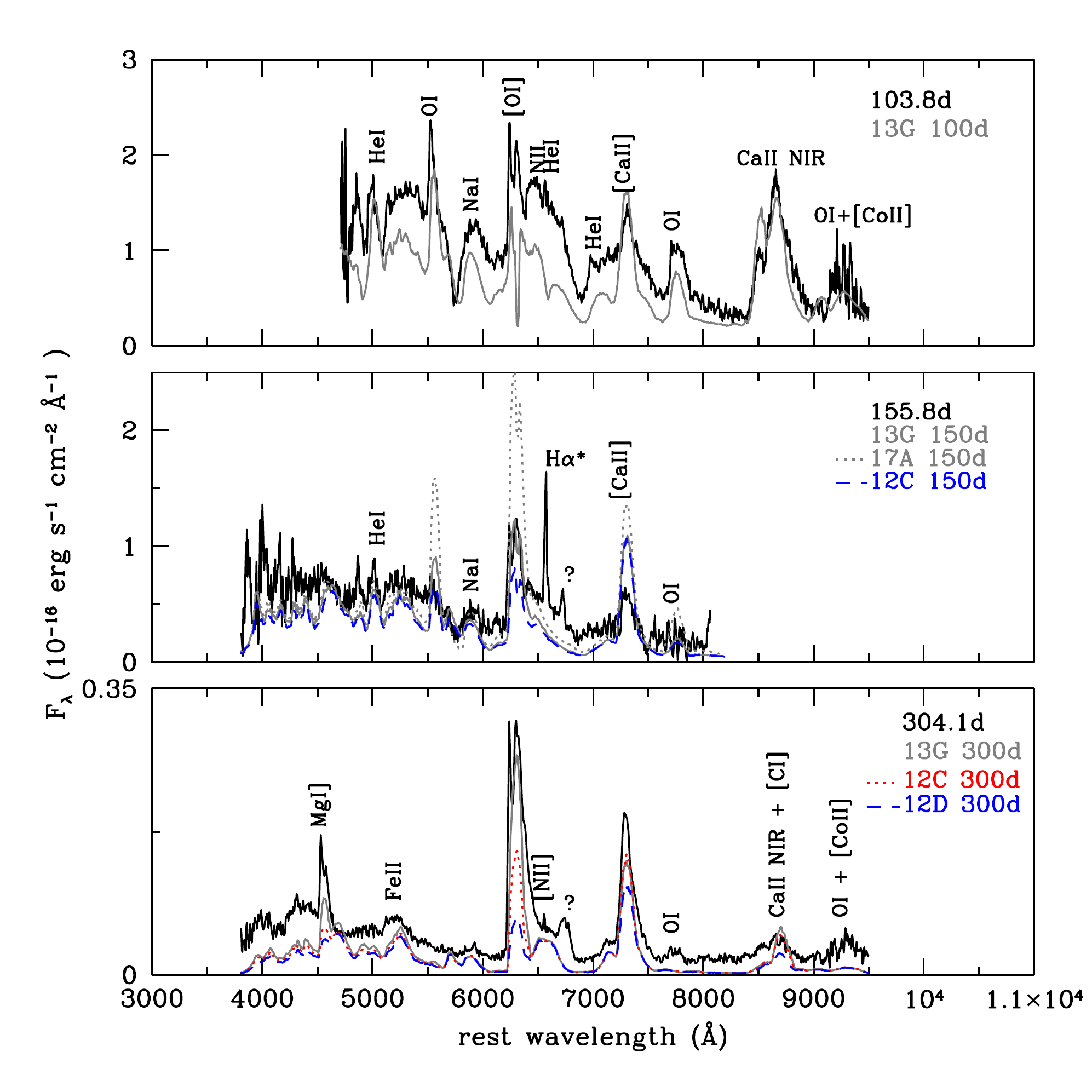}
 \caption{Late-time spectra of SN~2011fu compared to the best fitting spectral model 13G presented in \citealt{lateIIb}. In the middle panel we show our spectrum at $\sim156$~d and  the three models that vary only in progenitor mass. In the bottom panel  other, than  model 13G we show models 12C and 12D, which are the ones used to study molecule formation in the SN ejecta.  The spectral models have been scaled accounting for difference in phase  with respect to the SN spectra ($\Delta t$) by the factor $\rm exp (-2\Delta t/111.4)$, difference in $^{56}$Ni synthesised mass, and distance assumed for the models and the SN. }
 \label{fig:models_masses}
\end{figure*}

\begin{figure}
 \includegraphics[width=8cm]{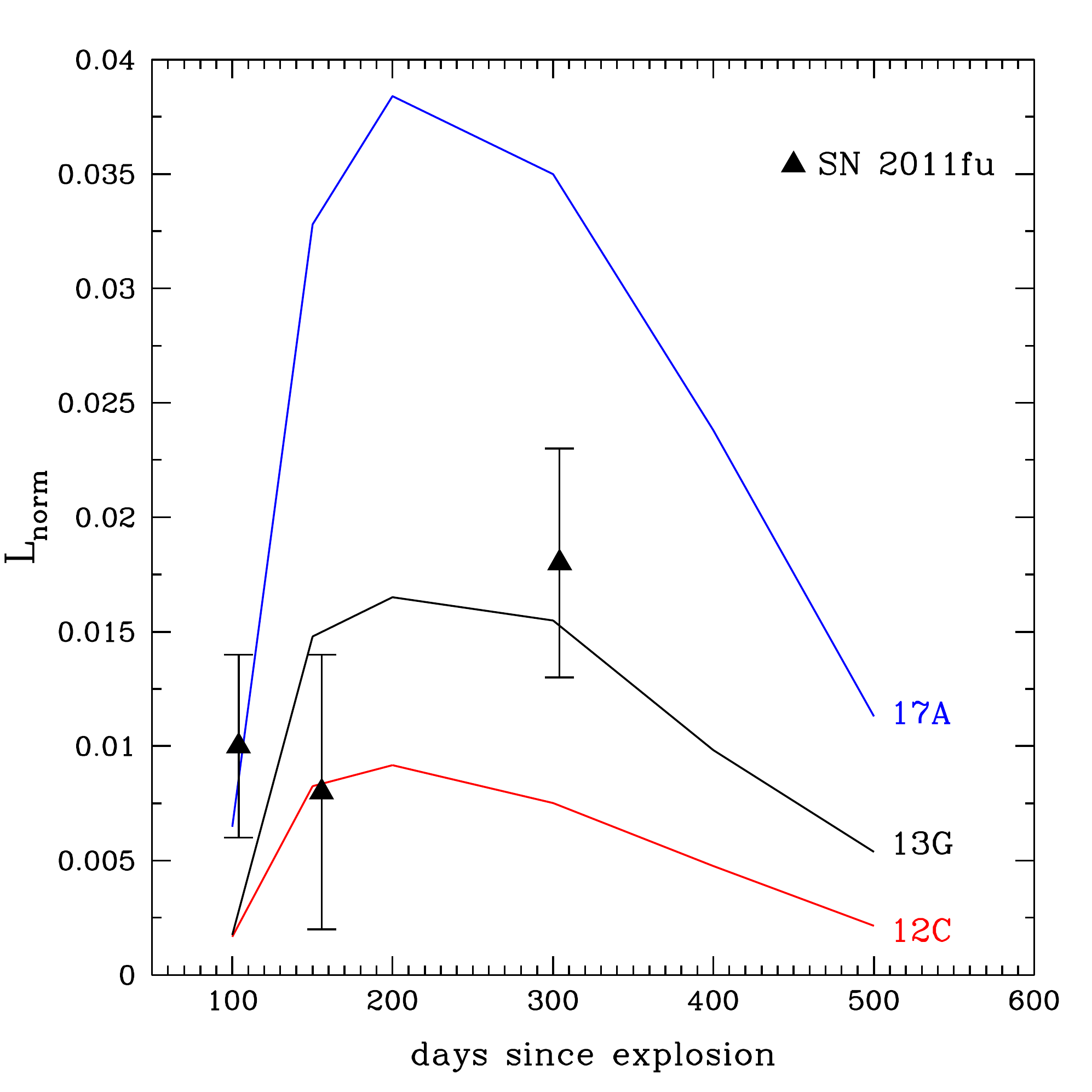}
 \caption{Late-time [O\,{\sc i}]~$\lambda \lambda$6300, 6364 luminosities of SN~2011fu (triangles) compared to model tracks 12C, 13G, and 17A of \citealt{lateIIb}. }
 \label{fig:oxygen_lumin}
\end{figure}

At all phases the overall best fitting model is 13G. The oxygen lines produced by the 17A model are more intense than those observed, while the 12C 
model underestimates the oxygen intensities. In Figure \ref{fig:models_masses} we present the three late-time SN spectra compared to the 13G spectral models at coeval phase. We have also indicated in the plot some 
of the most important features that arise in the late spectral modelling\footnote{H$\alpha$* is the narrow H$\alpha$ emission which we believe is associated with an H \,{\sc ii} region. The question mark at $\sim 6700$ marks the line we discussed in Section  \ref{specevol}.}. In the middle panel of the figure we have also plotted model 17A and 12C to reflect the discrepancy between the oxygen lines of the model and those observed of the spectrum. In Figure \ref{fig:oxygen_lumin} we present a comparison of the [O\,{\sc i}] $\lambda$$\lambda$ $6300$, $6364$ late-time luminosities of SN~20111fu and those corresponding to models 12C, 13G and 17A. 
All in all, these comparisons  indicate that the progenitor of SN~2011fu was likely at the lower end of the tested range, with  M$_{\rm ZAMS}\sim 13$M$_{\odot}$. 
The model that best fits the observed LC, however, corresponds to a progenitor star with a $5$~M$_{\odot}$ He core, i.e. M$_{\rm ZAMS}=18$~M$_{\odot}$. This value is higher but not inconsistent with the estimate we have obtained here. In fact, it is not surprising to have found different values for the progenitor's initial mass since the initial conditions and the methodology used in
both modelling methods are rather different. 
The hydrodynamical modelling of the LC uses stellar evolutionary models with different progenitor M$_{\rm ZAMS}$ and explodes these
structures for different values of explosion energies and nickel masses to compare with the data of SN~2011fu. 
The greatest degeneracy in the model is between mass and energy, this is the reason for modelling both the LC and the expansion velocities. Another source of uncertainty is introduced by the initial stellar evolutionary models. The code treats in a very approximate way the radiation transportation but correctly simulates the explosion dynamics.

The nebular spectral models use ejecta with nucleosynthesis
from the evolution and explosion of stars of different M$_{\rm ZAMS}$.
While the dynamic structure of the core is manually arranged to 
match the observed metal line widths in Type IIb SNe ($\sim3500$~km~s$^{-1}$)
as well as to capture effects of macroscopic mixing and clumping seen
in 2D simulations and inferred from observed line profiles,
the envelope follows a profile
obtained in a 1D explosion simulation by \cite{Melina12}.
Uncertainty is introduced by two sources; the dynamic structure of the ejecta, 
and the margin of error in the spectral modelling. Since the model reproduces well the metal
line profiles of SN~2011fu, the core velocity of 3500~km~s$^{-1}$  is likely accurate. The envelope absorbs quite
little of the gamma-ray energy (see Appendix A in \citealt{lateIIb}) so an
uncertainty in its density profile introduces only a moderate dispersion for
the flux levels. Thus, the model uncertainty is likely dominated by the error
in the $^{56}$Ni mixing as well as molecule and dust formation. The
$^{56}$Ni mixing used in 13G is chosen to match both the diffusion phase
LC and  the nebular spectra of SN~2011dh, but may be somewhat
different in SN~2011fu.

It would be interesting to analyse whether the
differences between the two modelling approaches persist if the post-explosion density structure of our
preferred LC model for SN~2011fu is used as initial condition of the spectral
modelling. However, this analysis is beyond the scope of this work. 
In any case, the mass estimates obtained from the two methods are 
consistent in the 
sense that we can discard the possibility of the progenitor being a single Wolf-Rayet star (with $M_{\mathrm{ZAMS}}\geq 25$).

To check for evidence of molecule formation in the ejecta of the SN, we compared models 12C (no molecular cooling; see table 4 of \citealt{lateIIb}) and 12D (with molecular cooling) with our late-time spectra. In the bottom panel of Figure \ref{fig:models_masses} the comparison at 300~d is shown. The oxygen lines are the ones that are most sensitive to molecule formation at 100 and 150 days along with [C\,{\sc i}]~$\lambda$8727 at 300~d. In our case, the model without molecular cooling is favoured, similarly to the case of SN~2011dh. We also looked at the contrast factor or oxygen zone density, i.e. the density ratio between the metal zones in the core and the Fe/Co/He 
zone, but the models with different oxygen zone densities do not show 
palpable
differences. However, we do note that the Mg\,{\sc i}] $\lambda$4571 line at 304~d is quite strong, which favours a high oxygen zone density, since the line is quite sensitive to this parameter, and this is in agreement with the results found for SN~2011dh in \cite{lateIIb}.

In summary, the comparisons of our nebular spectra with the models presented in \cite{lateIIb} point to a progenitor with an initial mass of $\sim13M_{\odot}$, likely no molecules were formed in the ejecta, and the oxygen zone density was possibly high.

\section{Conclusions}
\label{conclusion}
In this paper we have analysed optical and NIR data for the double-peaked type IIb SN~2011fu spanning from a few days to approximately 300 days after explosion. The double-peaked LC is brighter than those of other type IIb SNe such as SNe~1993J, 2008ax, 2011dh, and 2013df. It also presents a longer cooling phase after primary peak than SNe 1993J and 2013df. 

SN~2011fu's spectra evolve in a similar fashion to those of SN~1993J. During the first phases after explosion Balmer, He, Ca, and Fe lines are present in the spectra. At around 40~d, He lines dominate the spectra although H$\alpha$ is still present and persists up to later phases. Forbidden oxygen lines are clearly visible at $\sim62$~d, which indicates that the SN is entering the nebular phase. We derived H$\alpha$,  He\,{\sc i}~$\lambda5876$, and Fe\,{\sc ii}~$\lambda$$5169$ velocities from the absorption minima of the P-Cygni profiles in SN~2011fu's spectra, which are consistent with the results obtained for other type IIb SNe. At late times the most important features are [O\,{\sc i}]~$\lambda$$\lambda$ $6300$, $6364$, and  [Ca\,{\sc ii}]~$\lambda \lambda$$7291$, $7324$. In addition in our latest spectrum taken at 304~d,  Mg\,{\sc i}]~$\lambda$4571 and a line that could be due to [S\,{\sc ii}]~$\lambda\lambda$6713, 6730 are detected. 

 The nebular profiles of oxygen and magnesium indicate that the oxygen/magnesium zone is clumped. Calcium lines do not show similar profiles indicating, as expected, that they are formed at a different location.

Thanks to the hydrodynamical modelling of the pseudo-bolometric LC we conclude that SN~2011fu was the explosion of
an extended object (R~$\approx$~450~R$_\odot$), with 
a He core mass of 5~M$_\odot$ (M$_{\rm ZAMS}$~$\approx$~18M$_\odot$) and an ejecta mass of 3.5~M$_\odot$, an
explosion energy of 1.3 $\times 10^{51}$ erg, and a $^{56}$Ni 
mass of 0.15~M$_\odot$. In comparison with other SNe IIb, this object
seems to be a bit more massive, more energetic and to have produced more
$^{56}$Ni. The stellar envelope of its progenitor also seems to have been more massive than for other type IIb SNe ($\sim0.3$~M$_\odot$), implying the presence of more H in the SN ejecta.

We have compared our late-time spectra ($\geq$100~d) with late-time spectral models for type IIb SNe presented in \cite{lateIIb}. From this analysis, the overall best fitting value of the progenitor mass is 13~M$_\odot$,
similar to the M$_{\rm ZAMS}$  values that have previously been derived for SNe~1993J, 2008ax, and 2011dh (12-16~M$_{\rm ZAMS}$ , \citealt{lateIIb}). As for these other IIb SNe, we find also for SN~2011fu that molecule formation is non-existent. 

The extensive data-set of SN~2011fu presented in this paper have permitted us to derive some of the characteristics of the ejecta of one more type IIb SN  presenting a double-peaked LC, as well as to set constraints on some of its progenitor's properties.
SN~2011fu is another example of a type IIb SN which shows evidence that the progenitor mass was in the range 13-18~M$_\odot$. This is in the higher range of stars that are seen to explode \citep{SmarttPASA15}, which supports type IIb progenitor detections all being significantly more luminous than the II-P progenitors. Furthermore, as for all previously modelled IIb SNe, we determine a M$_{\rm ZAMS}$ value much lower that what would have been needed for a wind-stripped WR progenitor, and evidence is piling up that most, if not all, type IIb SNe come from binary stripped progenitors.

\section*{Acknowledgements}

AMG acknowledges financial support by the Spanish \textit{Ministerio de Econom\'ia y Competitividad} (MINECO) grant ESP2013-41268-R. NER acknowledges the support from the European Union Seventh Framework Programme (FP7/2007-2013) under grant agreement n. 267251 ``Astronomy Fellowships in Italy'' (AstroFIt).  ST acknowledges support by TRR33 ``The Dark Universe'' of the German Research Foundation (DFG). AP, SB, NER, and LT are partially supported by the PRIN-INAF 2014 within the project ``Transient Universe: unveiling new types of stellar explosions with PESSTO.''
This work was partly supported by the European Union FP7 programme through ERC grant number 320360.
SJS acknowledges funding from the European Research Council under the
European Union's Seventh Framework Programme (FP7/2007-2013)/ERC Grant
agreement n$^{\rm o}$ [291222] and STFC grants ST/I001123/1 and
ST/L000709/1.

Data of this work have been taken in the framework of the European supernova
collaboration involved in the ESO-NTT large programme 184.D-1140 led by Stefano Benetti.

This work is partially based on observations made with the Liverpool Telescope which is operated by Liverpool John Moores University with financial support from the UK Science and Technology Facilities Council, the Nordic Optical Telescope operated by the Nordic Optical Telescope Scientific Association, the \textit{Gran Telescopio de Canarias}, and the William Hershel Telescope (operated by the Isaac Newton Group) in the Spanish \textit{Observatorio del Roque de los Muchachos} (ORM) of the \textit{Instituto de Astrof\'isica de Canarias}; the 2.2m telescope of the \textit{Centro Astron\'omico Hispano Alem\'an} (Calar Alto, Spain); the 1.82-m Copernico Telescope operated by INAF - \textit{Osservatorio Astronomico di Padova} and the 1.22-m Galileo Telescope of \textit{Dipartimento di Fisica e Astonomia (Universit\'a di Padova)} at the Asiago Observatory (Italy); the \textit{Telescopio Carlos S\'anchez} operated by the \textit{Instituto de Astrof\'isica de Canarias} in the Spanish \textit{Observatorio del Teide};
 the Faulkes Telescope North situated on Mt. Haleakala (Hawaii, U.S.A.), and the 0.4-m telescope at the \textit{Observatorio Astron\'omico de Cantabria} (Spain).

We would like to thank F. Ciabattari and E. Mazzoni from the ISSP for providing us the discovery and confirmation image of the SN taken at the \textit{Osservatorio di Monte Agliale} (Lucca, Italy). 

This research has made use of the NASA/IPAC Extragalactic Database (NED) which is operated by 
the Jet Propulsion 
Laboratory, California Institute of Technology, under contract with the National Aeronautics and Space Administration.\\

%


\bigskip
\appendix
\null\newpage
\section{Instrumental Set-up}
\label{appendix}
\begin{table*}
\centering
\caption{Characteristics of the instruments used to acquire the photometric data of SN 2011fu.}
\begin{tabular}{cccccc} \hline
 Telescope & Instrument &    Field of View & Pixel Scale&  Filters & Key              \\
   &        &            &            ($\farcs$ pix$^{-1}$)      &      &                 \\

\hline
2-m Liverpool                           & RatCam           &  4'.6 x 4'.6      & 0.140          & uBVriz      & LT                     \\
1.82-m Asiago                        & AFOSC            &  8'.1 x 8'.1      & 0.473         & UBVRi       & A1p82                   \\
2.2-m Calar Alto                     & CAFOS            &  9' x 9'    & 0.530          &UBVRI        & Ca2p2                  \\
2.5-m Northern Optical                    & ALFOSC           &  6'.4 x 6'.4      & 0.190          &UBVRI        & NOT                    \\
4.2-m William Herschel                    & ACAM             &  9'.0 x 10'.4     & 0.250          &  griz       & WHT                     \\
2-m Faulkes North                       & fs02             &  10'.5 x 10'.5    & 0.304         &  BVRI       & FTN1                     \\
2-m Faulkes North                       & EM01             &   4'.7 x 4'.7     & 0.278         & BVRI        & FTN2                     \\
0.6-m Esteve Duran                  & ST-9XE           &  11'.7 x 11'.7    & 1.370          &   VI        &TED                       \\
0.4-m Cantabria                           & ST-8XE           &    24'.0 x 26'.0  &  0.900          &  VRI        &OAC                        \\
0.36-m Celestron C14                       & QHY9 CCD         &    25'.0 x 19'.0  & 0.450 &  unfiltered &Xing Gao                          \\
0.5-m Newtonian                     &  FLI Proline CCD &  20'.4 x 19'.8    & 2.320          &   unfiltered&F.Ciabattari/E.Mazzoni     \\                 
0.92-m Asiago Schmidt                      &  SCAM            &  48'.7 x 48'.7    & 1.437         &   BVRI      &  ASch                      \\    
10.4-m Gran Telescopio Canarias            & OSIRIS           &    7'.8 x 8'.5    & 0.127         &       gri &GTC        \\
1.52-m Carlos S\'anchez                    &   CAIN3            &   4'.2 x 4'.2     &  1.000            &  JHK$_{s}$  & TCS       \\

\hline
\end{tabular} \\ 

\begin{flushleft}

\end{flushleft}

\label{photometric_log}
\end{table*}

\begin{table*}
\centering
\caption{List of spectroscopic observations of SN 2011fu.}
\begin{tabular}{cccccccc} \hline
Date& Julian date & Phase$^{a}$ & Telescope+Instrument & Grism+Filter & Range &Resolving Power$^{b}$\\
    & (+2400000.00) & (d)&            &       & (\AA{})&             \\ 
\hline
2011/09/23 & 55828.3 & 3.8 &  AS1p82+AFOSC  & g4             &  3500--8450  &   300  \\
2011/09/24 & 55828.5 & 4.0 &   WHT+ISIS     &  R158R+R300B+gg495   &  3200--10000 &   1000,1300 \\
2011/09/25 & 55830.4 & 5.9 &  AS1p82+AFOSC  & g2+g4             & 3500--10200  &   200,300 \\
2011/09/26 & 55831.3 & 6.8 &  AS1p82+AFOSC  & g4             &  3500--8450  &   300 \\
2011/09/29 & 55833.6 & 9.1&  NOT+ALFOSC    & g4             &   3200--9100 &   500\\
2011/09/30 & 55835.3 & 10.8&   AS1p82+AFOSC & g2+g4             & 3500--10200  &   200,300 \\      
2011/10/01 & 55836.4 & 11.9&   AS1p82+AFOSC & g4             &  3500--8450  &   300 \\
2011/10/03 & 55838.5 & 14.0&   AS1p82+AFOSC & g4             &  3500--8450  &   300 \\
2011/10/06 & 55841.4 & 16.9 & Ca2p2+CAFOS   & B200+R200          & 3200--11000   &   500,500  \\
2011/10/10 & 55844.6 & 20.1 &  NOT+ALFOSC   & g4             &   3200--9100 &   500\\
2011/10/17 & 55851.5 & 27.0 & Ca2p2+CAFOS   & B200+R200      & 3200--11000   &   500,500  \\
2011/10/17 & 55851.6 & 27.1 & WHT+ACAM      & V400           &3500--9400  &   500\\
2011/10/24 & 55858.5 & 34.0 &  AS1p82+AFOSC & g4             &  3500--8450  &    300 \\ 
2011/10/27 & 55862.4 & 37.9 &  NOT+ALFOSC   & g4             &   3200--9100 &    500\\ 
2011/10/30 & 55864.6 & 40.1 &  AS1p82+AFOSC & g4             &  3500--8450  &   300 \\ 
2011/10/31 & 55866.4 & 41.9 &  AS1p82+AFOSC & g4             &  3500--8450  &   300 \\
2011/11/01 & 55867.4 & 42.9 & Ca2p2+CAFOS   &G200            &  3700--9500  &  500 \\
2011/11/07 & 55872.5 & 48.0 & WHT+ISIS      &R158R+R300B+gg495&  3200--10000  &   1000,1300 \\
2011/11/15 & 55881.4 & 56.9 & Ca2p2+CAFOS   & G200           &  3700--9500  &500 \\
2011/11/20 & 55886.4 & 61.9 &  AS1p82+AFOSC & g4             &  3500--8450  &  300 \\
2011/11/29 & 55894.5 & 70.0 & NOT+ALFOSC    & g4             &   3200--9100 & 500\\
2011/11/29 & 55895.4 & 70.9 & Ca2p2+CAFOS   & R200           & 6300--11000  & 500            \\
2011/12/17 & 55913.5 & 89.0 & TNG+DOLORES   & LRR+LRB        &  3000-10000            & 460,460 \\
2011/12/20 & 55916.4 & 91.9 & AS1p82+AFOSC  & g4             &  3500--8450  &  300 \\
2011/12/21 & 55917.4 & 92.9 &  NOT+ALFOSC   & g4+g5          &   3200--10250 & 500,500\\
2012/01/01 & 55928.4 & 103.9& Ca2p2+CAFOS   & G200+gg495     & 4800--9500  & 500            \\
2012/02/22 & 55980.3 & 155.8& AS1p82+AFOSC  & g4             &  3500--8450  & 300 \\
2012/07/20 & 56128.6 & 304.1& GTC+OSIRIS    & R300B+R300R    & 3600--10000   & 430 \\
\hline
\end{tabular} \\
\flushleft $^{a}$ Phase with respect to our adopted explosion epoch $\rm JD =2455824.5\pm0.7$.\\
\flushleft$^{b}$ The resolution was measured from the full width at half maximum of the night sky lines. The resolving power values reported here are for a reference wavelength of 6500~\AA{}. 
\label{spectroscopic_log}
\end{table*}

\null\newpage
\section{SN~2011fu's photometric data}

\begin{table*}
\centering
\caption{Optical Bessell  photometry of SN~2011fu. The \textit{U} data are not S-corrected,  note that the reported \textit{RI} LT data are the S-corrected Sloan \textit{ri} magnitudes.}
\begin{tabular}{ccccccccc} \hline 
    Date & JD & Phase$^{a}$    &    \textit{U}  &  \textit{B} & \textit{V}                & \textit{R} & \textit{I} & Key\\ 
         & (+2400000.00)   & (d) & (mag)& (mag)&(mag)&(mag)&(mag)& \\
  \hline
 
 2011/09/20  &55825.2   & 0.7  &$	-	$&$	-	$&$	16.32	\pm	0.14	$&$	-	$&$	-	$&Xing Gao\\
 2011/09/21  &55825.5   & 1.0  &$	-	$&$	-	$&$	-	$&$	16.07	\pm	0.13	$&$	-	$&F. Ciabattari/E.Mazzoni\\
 2011/09/21  &55826.4   & 1.9  &$	-	$&$	-	$&$	16.21	\pm	0.15	$&$	-	$&$	-	$&Xing Gao\\
 2011/09/21  &55826.4   & 1.9  &$	-	$&$	-	$&$	-	$&$	16.01	\pm	0.08	$&$	-	$&F. Ciabattari/E.Mazzoni\\
 2011/09/23  &55828.3   & 3.8  &$	-	$&$	-	$&$	-	$&$	16.27	\pm	0.02	$&$	-	$&A1p82\\
 2011/09/23  &55828.4   & 3.9  &$	-	$&$	-	$&$	16.45	\pm	0.24	$&$	-	$&$	-	$&Xing Gao\\
 2011/09/25  &55830.5   & 6.0  &$	16.52	\pm	0.02	$&$	17.10	\pm	0.02	$&$	16.83	\pm	0.01	$&$	16.65	\pm	0.02	$&$	16.52	\pm	0.02	$&A1p82\\
 2011/09/26  &55831.2   & 6.7  &$	-	$&$	-	$&$	17.09	\pm	0.15	$&$	-	$&$	-	$&Xing Gao\\
 2011/09/26  &55831.4   & 6.9  &$	16.43	\pm	0.03	$&$	17.31	\pm	0.02	$&$	17.02	\pm	0.01	$&$	16.80	\pm	0.02	$&$	16.62	\pm	0.02	$&A1p82\\
 2011/09/27  &55831.5   & 7.0  &$	-	$&$	17.40	\pm	0.02	$&$	17.19	\pm	0.01	$&$	16.78	\pm	0.04	$&$	16.92	\pm	0.03	$&LT\\
 2011/09/29  & 55833.5   &9.0 &$	-	$&$	17.85	\pm	0.02	$&$	17.41	\pm	0.03	$&$	17.12	\pm	0.04	$&$	16.84	\pm	0.03	$&LT\\
 2011/09/29  & 55834.5   &10.0 &$	17.86	\pm	0.03	$&$	17.93	\pm	0.02	$&$	17.57	\pm	0.01	$&$	17.28	\pm	0.02	$&$	17.00	\pm	0.02	$&A1p82\\
 2011/09/29  & 55834.6   &10.1 &$	-	$&$	17.91	\pm	0.02	$&$	17.55	\pm	0.02	$&$	17.11	\pm	0.02	$&$	16.87	\pm	0.03	$&LT\\
 2011/09/30  & 55835.3   &10.8 &$	-	$&$	-	$&$	17.40	\pm	0.14	$&$	-	$&$	-	$&Xing Gao\\
 2011/09/30  & 55835.4   &10.9 &$	-	$&$	17.97	\pm	0.02	$&$	17.55	\pm	0.02	$&$	17.24	\pm	0.01	$&$	16.98	\pm	0.05	$&A1p82\\
 2011/10/01  & 55835.5   &11.0 &$	-	$&$	17.95	\pm	0.01	$&$	17.55	\pm	0.03	$&$	17.08	\pm	0.03	$&$	17.02	\pm	0.03	$&LT\\
 2011/10/01  & 55836.4   &11.9 &$	17.74	\pm	0.06	$&$	17.96	\pm	0.02	$&$	17.51	\pm	0.01	$&$	17.19	\pm	0.01	$&$	17.02	\pm	0.03	$&A1p82\\
 2011/10/03  & 55837.5   &13.0 &$	-	$&$	17.91	\pm	0.02	$&$	17.46	\pm	0.03	$&$	17.07	\pm	0.43	$&$	16.83	\pm	0.05	$&LT\\
 2011/10/03  & 55838.5   &14.0 &$	17.96	\pm	0.06	$&$	17.84	\pm	0.02	$&$	17.37	\pm	0.01	$&$	17.06	\pm	0.01	$&$	16.90	\pm	0.03	$&A1p82\\
 2011/10/05  & 55840.2   &15.7 &$	-	$&$	-	$&$	17.34	\pm	0.18	$&$	-	$&$	-	$&Xing Gao\\
 2011/10/06  & 55840.5   &16.0 &$	-	$&$	17.79	\pm	0.02	$&$	17.26	\pm	0.02	$&$	16.91	\pm	0.03	$&$	16.64	\pm	0.04	$&LT\\
 2011/10/06  & 55841.4   &16.9 &$	17.45	\pm	0.04	$&$	17.64	\pm	0.03	$&$	17.17	\pm	0.02	$&$	16.78	\pm	0.02	$&$	16.71	\pm	0.06	$&Ca2p2\\
 2011/10/07  & 55841.5   &17.0 &$	-	$&$	17.60	\pm	0.03	$&$	17.24	\pm	0.05	$&$	16.77	\pm	0.06	$&$	16.54	\pm	0.05	$&LT\\
 2011/10/08  & 55842.5   &18.0 &$	-	$&$	-	$&$	17.08	\pm	0.02	$&$	-	$&$	-	$&LT\\
 2011/10/08  & 55843.3   &18.8 &$	-	$&$	-	$&$	16.96	\pm	0.15	$&$	-	$&$	-	$&Xing Gao\\
 2011/10/10  & 55844.5   &20.0 &$	-	$&$	17.58	\pm	0.02	$&$	16.99	\pm	0.03	$&$	16.64	\pm	0.03	$&$	16.46	\pm	0.03	$&LT\\
 2011/10/10  & 55844.5   &20.0 &$	-	$&$	-	$&$	16.90	\pm	0.06	$&$	-	$&$	16.66	\pm	0.05	$&TED\\
 2011/10/11  & 55845.5   &21.0 &$	-	$&$	17.60	\pm	0.04	$&$	16.97	\pm	0.03	$&$	16.61	\pm	0.07	$&$	16.40	\pm	0.05	$&LT\\
 2011/10/11  & 55845.5   &21.0 &$	-	$&$	-	$&$	16.94	\pm	0.04	$&$	16.62	\pm	0.03	$&$	-	$&OAC\\
 2011/10/11  & 55845.5   &21.0 &$	-	$&$	-	$&$	16.94	\pm	0.02	$&$	-	$&$	16.56	\pm	0.03	$&TED\\
 2011/10/12  & 55846.5   &22.0 &$	-	$&$	17.47	\pm	0.03	$&$	16.92	\pm	0.04	$&$	16.57	\pm	0.03	$&$	16.39	\pm	0.03	$&LT\\
 2011/10/12  & 55846.5   &22.0 &$	-	$&$	-	$&$	-	$&$	16.58	\pm	0.05	$&$	16.50	\pm	0.06	$&OAC\\
 2011/10/13  & 55847.5   &23.0 &$	-	$&$	17.53	\pm	0.04	$&$	16.90	\pm	0.04	$&$	16.54	\pm	0.04	$&$	16.45	\pm	0.03	$&LT\\
 2011/10/14  & 55848.5   &24.0 &$	-	$&$	17.58	\pm	0.07	$&$	16.92	\pm	0.06	$&$	16.52	\pm	0.06	$&$	16.36	\pm	0.04	$&LT\\
 2011/10/15  & 55849.5   &25.0 &$	-	$&$	17.73	\pm	0.03	$&$	16.96	\pm	0.03	$&$	16.54	\pm	0.06	$&$	16.37	\pm	0.04	$&LT\\
 2011/10/16  & 55851.4   &26.9 &$	-	$&$	-	$&$	17.07	\pm	0.07	$&$	-	$&$	16.56	\pm	0.04	$&TED\\
 2011/10/16  & 55851.5   &27.0 &$	18.12	\pm	0.10	$&$	17.87	\pm	0.03	$&$	17.08	\pm	0.02	$&$	16.66	\pm	0.02	$&$	16.56	\pm	0.04	$&Ca2p2\\
 2011/10/18  & 55852.5   &28.0 &$	-	$&$	18.05	\pm	0.02	$&$	17.14	\pm	0.02	$&$	16.63	\pm	0.03	$&$	16.44	\pm	0.03	$&LT\\
 2011/10/18  & 55853.2   &28.7 &$	-	$&$	-	$&$	17.30	\pm	0.13	$&$	-	$&$	-	$&Xing Gao\\
 2011/10/18  & 55853.4   &28.9 &$	-	$&$	-	$&$	17.24	\pm	0.07	$&$	-	$&$	16.53	\pm	0.04	$&TED\\
 2011/10/22  & 55857.2   &32.7 &$	-	$&$	-	$&$	17.43	\pm	0.12	$&$	-	$&$	-	$&Xing Gao\\
 2011/10/23  & 55857.6   &33.1 &$	-	$&$	18.72	\pm	0.03	$&$	17.57	\pm	0.03	$&$	16.83	\pm	0.06	$&$	16.59	\pm	0.04	$&LT\\
 2011/10/23  & 55858.2   &33.7 &$	-	$&$	-	$&$	17.41	\pm	0.14	$&$	-	$&$	-	$&Xing Gao\\
 2011/10/24  & 55858.6   &34.1 &$	-	$&$	18.70	\pm	0.02	$&$	17.61	\pm	0.01	$&$	16.98	\pm	0.01	$&$	16.73	\pm	0.03	$&A1p82\\
 2011/10/24  & 55859.2   &34.7 &$	-	$&$	-	$&$	17.47	\pm	0.13	$&$	-	$&$	-	$&Xing Gao\\
 2011/10/25  & 55859.5   &35.0 &$	-	$&$	18.78	\pm	0.04	$&$	17.63	\pm	0.03	$&$	16.92	\pm	0.06	$&$	16.70	\pm	0.04	$&LT\\
 2011/10/25  & 55860.4   &35.9 &$	-	$&$	-	$&$	17.79	\pm	0.05	$&$	-	$&$	16.89	\pm	0.04	$&TED\\
 2011/10/26  & 55860.5   &36.0 &$	-	$&$	18.81	\pm	0.04	$&$	17.68	\pm	0.03	$&$	16.97	\pm	0.06	$&$	16.67	\pm	0.04	$&LT\\
 2011/10/27  & 55861.5   &37.0 &$	-	$&$	18.91	\pm	0.04	$&$	17.79	\pm	0.03	$&$	16.99	\pm	0.06	$&$	17.08	\pm	0.04	$&LT\\
 2011/10/28  & 55862.5   &38.0 &$	19.70	\pm	0.05	$&$	19.10	\pm	0.03	$&$	17.77	\pm	0.02	$&$	17.13	\pm	0.02	$&$	16.64	\pm	0.01	$&NOT\\
 2011/10/28  & 55862.5   &38.0 &$	-	$&$	19.08	\pm	0.04	$&$	17.80	\pm	0.03	$&$	17.21	\pm	0.02	$&$	16.76	\pm	0.02	$&LT\\
 2011/10/29  & 55863.5   &39.0 &$	-	$&$	19.06	\pm	0.04	$&$	17.83	\pm	0.03	$&$	17.14	\pm	0.06	$&$	16.74	\pm	0.04	$&LT\\

 \hline
\end{tabular} \\

\label{UBVRI_phot}
\end{table*}

\begin{table*}
\centering
\contcaption{}
\begin{tabular}{ccccccccc} \hline 
    Date & JD & Phase$^{a}$    &    \textit{U}  &  \textit{B} & \textit{V}                & \textit{R} & \textit{I} & Key\\ 
         & (+2400000.00)   & (d) & (mag)& (mag)&(mag)&(mag)&(mag)& \\
  \hline
 2011/10/30  & 55864.5   &40.0 &$	-	$&$	19.07	\pm	0.07	$&$	17.80	\pm	0.05	$&$	17.26	\pm	0.06	$&$	-	$&LT\\
 2011/10/30  & 55864.5   &40.0 &$	-	$&$	18.97	\pm	0.06	$&$	17.89	\pm	0.03	$&$	17.20	\pm	0.03	$&$	17.01	\pm	0.03	$&A1p82\\
 2011/10/31  & 55866.5   &42.0 &$	-	$&$	19.10	\pm	0.06	$&$	17.97	\pm	0.03	$&$	17.31	\pm	0.02	$&$	17.03	\pm	0.03	$&A1p82\\
 2011/11/01  & 55866.6   &42.1 &$	-	$&$	19.17	\pm	0.06	$&$	17.99	\pm	0.05	$&$	17.20	\pm	0.03	$&$	16.90	\pm	0.02	$&LT\\
2011/11/01  & 55867.4   &42.9 &$	20.02	\pm	0.10	$&$	19.06	\pm	0.02	$&$	18.03	\pm	0.02	$&$	17.23	\pm	0.02	$&$	17.12	\pm	0.05	$&Ca2p2\\
 2011/11/05  & 55871.1   &46.6 &$	-	$&$	-	$&$	17.88	\pm	0.17	$&$	-	$&$	-	$&Xing Gao\\
 2011/11/06  & 55872.1   &47.6 &$	-	$&$	-	$&$	18.00	\pm	0.15	$&$	-	$&$	-	$&Xing Gao\\
 2011/11/07  & 55872.6   &48.1 &$	-	$&$	-	$&$	-	$&$	17.57	\pm	0.18	$&$	17.15	\pm	0.03	$&WHT\\
 2011/11/11  &55876.9  & 52.4  &$	-	$&$	-	$&$	-	$&$	17.59	\pm	0.03	$&$	17.30	\pm	0.03	$&FTN1\\
 2011/11/14  &55880.2  & 55.7  &$	-	$&$	19.41	\pm	0.10	$&$	18.27	\pm	0.10	$&$	17.42	\pm	0.09	$&$	17.32	\pm	0.09	$&ASch\\
 2011/11/15  &55881.4  & 56.9  &$	-	$&$	19.29	\pm	0.04	$&$	18.25	\pm	0.03	$&$	17.65	\pm	0.02	$&$	17.42	\pm	0.02	$&Ca2p2\\
 2011/11/17  &55882.9  &48.4  &$	-	$&$	19.51	\pm	0.06	$&$	18.28	\pm	0.05	$&$	17.78	\pm	0.03	$&$	17.43	\pm	0.03	$&FTN1\\
 2011/11/19  &55885.4  & 60.9  &$	-	$&$	19.49	\pm	0.04	$&$	18.49	\pm	0.02	$&$	17.82	\pm	0.02	$&$	17.45	\pm	0.03	$&A1p82\\
 2011/11/21  &55886.5  & 62.0  &$	-	$&$	19.64	\pm	0.05	$&$	18.56	\pm	0.03	$&$	17.76	\pm	0.04	$&$	17.39	\pm	0.02	$&A1p82\\
 2011/11/22  &55888.1  & 63.6  &$	-	$&$	-	$&$	18.38	\pm	0.13	$&$	-	$&$	-	$&Xing Gao\\
 2011/11/23  &55888.5  & 64.0  &$	-	$&$	19.47	\pm	0.05	$&$	18.40	\pm	0.06	$&$	17.67	\pm	0.04	$&$	17.42	\pm	0.05	$&LT\\
 2011/11/25  &55890.8  & 66.3  &$	-	$&$	19.66	\pm	0.06	$&$	18.45	\pm	0.05	$&$	17.80	\pm	0.03	$&$	17.57	\pm	0.03	$&FTN1\\
 2011/11/27  &55893.0  & 68.5  &$	-	$&$	-	$&$	18.41	\pm	0.11	$&$	17.78	\pm	0.13	$&$	17.59	\pm	0.13	$&FTN1\\
 2011/11/29  &55894.9  & 70.4  &$	-	$&$	19.52	\pm	0.07	$&$	18.50	\pm	0.05	$&$	17.94	\pm	0.04	$&$	17.66	\pm	0.03	$&FTN1\\
 2011/11/29  &55894.9  & 70.4  &$	-	$&$	19.68	\pm	0.03	$&$	18.59	\pm	0.02	$&$	17.96	\pm	0.02	$&$	17.46	\pm	0.01	$&NOT\\
 2011/11/30  &55896.4  & 71.9  &$	-	$&$	19.63	\pm	0.05	$&$	18.57	\pm	0.06	$&$	17.89	\pm	0.05	$&$	17.55	\pm	0.06	$&LT\\
 2011/12/01  &55896.8  & 72.3  &$	-	$&$	-	$&$	18.53	\pm	0.05	$&$	17.97	\pm	0.04	$&$	17.71	\pm	0.03	$&FTN1\\
 2011/12/06  &55901.8  & 77.3  &$	-	$&$	19.91	\pm	0.06	$&$	18.56	\pm	0.05	$&$	18.04	\pm	0.03	$&$	17.77	\pm	0.03	$&FTN2\\
 2011/12/12  &55908.4  & 83.9  &$	-	$&$	-	$&$	18.87	\pm	0.07	$&$	18.09	\pm	0.05	$&$	17.73	\pm	0.01	$&LT\\
 2011/12/20  &55916.4  & 91.9  &$	-	$&$	19.71	\pm	0.09	$&$	18.94	\pm	0.09	$&$	18.34	\pm	0.04	$&$	17.97	\pm	0.05	$&A1p82\\
 2011/12/22  &55917.8  & 93.3  &$	20.01	\pm	0.07	$&$	19.80	\pm	0.03	$&$	18.88	\pm	0.03	$&$	18.38	\pm	0.03	$&$	17.96	\pm	0.06	$&NOT\\
 2011/12/22  &55917.8  & 93.3  &$	-	$&$	19.75	\pm	0.06	$&$	18.85	\pm	0.05	$&$	18.40	\pm	0.03	$&$	18.11	\pm	0.03	$&FTN2\\
 2012/01/01  & 55928.3  &103.8  &$	-	$&$	19.80	\pm	0.02	$&$	19.07	\pm	0.03	$&$	18.52	\pm	0.03	$&$	18.34	\pm	0.05	$&Ca2p2\\
 2012/01/09  & 55936.5  &112.0  &$	-	$&$	19.96	\pm	0.08	$&$	19.18	\pm	0.07	$&$	18.72	\pm	0.05	$&$	18.44	\pm	0.05	$&LT\\
 2012/01/12  & 55939.4  &114.9  &$	-	$&$	19.92	\pm	0.03	$&$	19.33	\pm	0.03	$&$	18.85	\pm	0.02	$&$	18.67	\pm	0.05	$&Ca2p2\\
 2012/01/15  & 55942.3  &117.8  &$	-	$&$	20.00	\pm	0.08	$&$	19.24	\pm	0.07	$&$	18.76	\pm	0.06	$&$	18.41	\pm	0.06	$&LT\\
 2012/01/27  & 55954.4  &129.9  &$	-	$&$	20.32	\pm	0.08	$&$	19.48	\pm	0.07	$&$	19.04	\pm	0.06	$&$	18.80	\pm	0.04	$&LT\\
 2012/02/17  & 55975.3  &150.8  &$	-	$&$	20.55	\pm	0.13	$&$	19.98	\pm	0.09	$&$	19.60	\pm	0.06	$&$	19.19	\pm	0.05	$&A1p82\\
 2012/02/22  & 55980.3  &155.8  &$	-	$&$	20.69	\pm	0.18	$&$	19.99	\pm	0.15	$&$	19.61	\pm	0.04	$&$	19.01	\pm	0.04	$&A1p82\\
 2012/02/23  & 55981.4  &156.9  &$	-	$&$	-	$&$	20.13	\pm	0.08	$&$	19.73	\pm	0.03	$&$	19.27	\pm	0.08	$&LT\\
 2012/02/25  & 55983.4  &148.9  &$	-	$&$	-	$&$	19.89	\pm	0.25	$&$	19.48	\pm	0.11	$&$	19.24	\pm	0.13	$&LT\\
 2012/07/20  & 56128.6 & 304.1 &$	-	$&$	-	$&$	-	$&$	21.54	\pm	0.02	$&$	21.45	\pm	0.04	$&GTC\\

 \hline
\end{tabular} \\
\begin{flushleft}
$^{a}${Phase in days with respect to the adopted explosion date JD = $2455824.5\pm0.7$}\\

\end{flushleft}

\end{table*}

 \begin{table*}
\centering
\caption{Optical Sloan photometry of SN~2011fu in the Vega system.}
\begin{tabular}{ccccccc} \hline 
    Date & JD & Phase$^{a}$    &    \textit{u}  &  \textit{g} & \textit{z}   & Key \\ 
         & (+2400000.00) & (d) & (mag)& (mag)&(mag)&\\
 \hline        
2011/09/27	&   55831.5	&	7.0       &$	16.43	\pm	0.06	$&$	-	$&$16.37\pm	0.03	$&	LT	\\
2011/09/29	&   55833.5	&	9.0      &$	17.02\pm	0.04	$&$	-       $&$16.61\pm	0.04	$&	LT	\\
2011/09/30	&   55834.6	&	10.1      &$	17.37	\pm	0.06	$&$	-	$&$16.77\pm	0.03	$&	LT	\\
2011/10/01	&   55835.5	&	11.0      &$	17.45	\pm	0.09	$&$	-	$&$16.70\pm	0.03	$&	LT	\\
2011/10/03	&   55837.5	&	13.0      &$	-	$&$	-	                $&$16.66\pm	0.05	$&	LT	\\
2011/10/06	&   55840.5	&	16.0      &$	17.38	\pm	0.04	$&$	-	$&$16.49\pm	0.04	$&	LT	\\
2011/10/08	&   55842.5	&	18.0      &$	-	$&$	-	                $&$16.40\pm	0.06	$&	LT	\\
2011/10/10	&   55844.5	&	20.0      &$	17.23	\pm	0.03	$&$	-	$&$16.32\pm	0.03	$&	LT	\\
2011/10/11	&   55845.5	&	21.0      &$	17.5	\pm	0.13	$&$	-	$&$16.28\pm	0.05	$&	LT	\\
2011/10/12	&   55846.5	&	22.0      &$	17.71	\pm	0.11	$&$	-	$&$16.30\pm	0.03	$&	LT	\\
2011/10/13	&   55847.5	&	25.0      &$	17.75	\pm	0.03	$&$	-	$&$16.34\pm	0.05	$&	LT	\\
2011/10/14	&   55848.5	&	24.0      &$	17.74	\pm	0.11	$&$	-	$&$16.36\pm	0.05	$&	LT	\\
2011/10/15	&   55849.5	&	25.0      &$	17.82	\pm	0.10	$&$	-	$&$16.22\pm	0.05	$&	LT	\\
2011/10/18	&   55852.5	&	28.0      &$	18.22	\pm	0.10	$&$	-	$&$16.32\pm	0.03	$&	LT	\\
2011/10/23	&   55857.6	&	33.1      &$	-	$&$	-	                $&$16.46\pm	0.05	$&	LT	\\
2011/10/25	&   55859.5	&	35.0      &$	-	$&$	-	                $&$16.54\pm	0.05	$&	LT	\\
2011/10/26	&   55860.5	&	36.0      &$	-	$&$	-	                $&$16.52\pm	0.05	$&	LT	\\
2011/10/27	&   55861.5	&	37.0      &$	-	$&$	-	                $&$16.59\pm	0.05	$&	LT	\\
2011/10/28	&   55862.5	&	38.0      &$	19.37	\pm	0.03	$&$	-	$&$16.66\pm	0.02	$&	LT	\\
2011/10/29	&   55863.5	&	39.0      &$	19.24	\pm	0.07	$&$	-	$&$16.65\pm	0.04	$&	LT	\\
2011/10/30	&   55864.5	&	40.0      &$	-	$&$	-	                $&$16.75\pm	0.04	$&	LT	\\
2011/11/01	&   55866.6	&	42.1      &$	-	$&$	-                       $&$16.69\pm	0.02	$&	LT	\\
2011/11/07	 &    55872.6	&	48.1      &$	-	$&$	18.72	\pm	0.29	$&$16.85\pm	0.05	$&	WHT	\\
2011/11/23	&   55888.5	&	64.0      &$	-	$&$	-	                $&$17.10\pm	0.06	$&	LT	\\
2011/11/30	&   55896.4	&	71.9      &$	-	$&$	-	                $&$17.35\pm	0.05	$&	LT	\\
2011/12/12	&   55908.4	&	83.9      &$	-	$&$	-	                $&$17.51\pm	0.05	$&	LT	\\
2012/01/09	&   55936.5	&	112.0     &$	-	$&$	-	                $&$18.07\pm	0.07	$&	LT	\\
2012/01/15	&   55942.3	&	117.8     &$	-	$&$	-	                $&$18.14\pm	0.08	$&	LT	\\
2012/01/27	&   55954.4	&	129.9     &$	-	$&$	-	                $&$18.25\pm	0.07	$&	LT	\\
2012/02/23	&   55981.4	&	156.9     &$	-       $&$	-	                $&$18.82\pm	0.16	$&	LT	\\
2012/07/20	& 56128.6	&	304.1     &$	-	$&$	22.41	\pm	0.15	$&$	-	$&	GTC	\\

 \hline
\end{tabular} \\
\begin{flushleft}
$^{a}${Phase in days with respect to the adopted explosion date JD = $2455824.5\pm0.7$}\\

\end{flushleft}

\label{ugz_phot}
\end{table*}
        
  \begin{table*}
\centering
\caption{NIR photometry of SN~2011fu.}
\begin{tabular}{ccccccc} \hline 
    Date & JD & Phase$^{a}$    &    \textit{J}  &  \textit{H} & \textit{K$_{\rm s}$}   & Key \\ 
         & (+2400000.00) & (d) & (mag)& (mag)&(mag)&\\  
\hline         
2011/10/06  & 55840.6&16.1	&$	16.20	\pm	0.12	$&$	16.10	\pm	0.14	$&$	15.75	\pm	0.14	$& 	TCS \\
2011/10/10  & 55844.6&20.1	&$	16.08	\pm	0.12	$&$	-	$&$	15.72	\pm	0.27	$& 	TCS \\
2011/10/12  & 55847.4&23.0	&$	16.15	\pm	0.08	$&$	15.70	\pm	0.15	$&$	15.63	\pm	0.34	$& 	TCS \\
2011/10/29  & 55864.5&40.0	&$	16.28	\pm	0.13	$&$	-	$&$	-	$& 	TCS \\
2011/10/31  & 55866.5&42.0	&$	16.54	\pm	0.15	$&$	16.06	\pm	0.15	$&$	15.77	\pm	0.21	$& 	TCS \\
2011/11/04  & 55869.5&45.0	&$	16.74	\pm	0.15	$&$	16.27	\pm	0.12	$&$	16.17	\pm	0.27	$& 	TCS \\

  \hline
\end{tabular} \\
\begin{flushleft}
$^{a}${Phase in days with respect to the adopted explosion date JD = $2455824.5\pm0.7$}\\

\end{flushleft}

\label{jhk_phot}
\end{table*}

 \begin{table*}
\centering
\caption{Johnson-Cousins optical and 2MASS NIR magnitudes and associated errors for the stellar sequence used in the calibration process of SN~2011fu's photometry.}
\begin{tabular}{ccccccccc} \hline 
    Star    &    \textit{U}  &  \textit{B} & \textit{V}                & \textit{R} & \textit{I} & \textit{J}  &  \textit{H} & \textit{K$_{\rm s}$}    \\ 
            & (mag) & (mag) & (mag) & (mag)& (mag) & (mag) & (mag)& (mag) \\
  \hline
1 &  $19.26\pm0.10$   & $18.66\pm0.02$ &$17.74\pm0.02$ &$17.20\pm0.03$ &$16.65\pm0.03$ &-- &--&--\\
2 &  $18.25\pm0.10$   & $18.17\pm0.02$ &$17.52\pm0.02$ &$17.10\pm0.05$ &$16.79\pm0.03$ &--&--&-- \\
3 &  $18.26\pm0.06$   & $18.25\pm0.02$ &$17.55\pm0.02$ &$17.24\pm0.02$ &$16.87\pm0.03$ & $16.24\pm0.12$ &$15.96 \pm0.19  $&$15.79 \pm0.26 $\\ 
4 &  $18.98\pm0.09$   & $18.14\pm0.02$ &$17.18\pm0.02$ &$16.63\pm0.02$ &$16.14\pm0.03$ & $15.45\pm0.06$ &$14.81 \pm0.08  $&$14.81 \pm0.11 $\\
5 &  $17.75\pm0.04$   & $17.83\pm0.02$ &$17.25\pm0.02$ &$16.90\pm0.02$ &$16.58\pm0.02$ & $15.91\pm0.08$ &$15.77 \pm0.15  $&$15.51 \pm0.20 $\\
6 &  $15.68\pm0.09$   & $15.51\pm0.02$ &$14.88\pm0.07$ &$14.47\pm0.07$ &$14.11\pm0.06$ & $13.49\pm0.02$ &$13.16 \pm0.03  $&$13.07 \pm0.03 $\\
7 &  $17.88\pm0.04$   & $17.77\pm0.02$ &$17.14\pm0.03$ &$16.77\pm0.03$ &$16.41\pm0.02$ & $15.83\pm0.08$ &$15.30 \pm0.10  $&$15.42 \pm0.19 $\\
8 &  $19.94\pm0.05$   & $18.82\pm0.02$ &$17.58\pm0.03$ &$16.82\pm0.02$ &$16.15\pm0.03$ & $15.11\pm0.05$ &$14.54 \pm0.05  $&$14.28 \pm0.07 $\\
9 &  $17.60\pm0.04$   & $17.61\pm0.02$ &$17.07\pm0.02$ &$16.75\pm0.03$ &$16.42\pm0.02$ & $15.97\pm0.09$ &$15.73 \pm0.15  $&$15.25  $*\\ 
10&  $19.48\pm0.10$   & $18.89\pm0.02$ &$18.02\pm0.02$ &$17.57\pm0.02$ &$17.14\pm0.02$  &--&--&--\\
11&  $15.21\pm0.04$   & $15.21\pm0.02$ &$14.71\pm0.02$ &$14.37\pm0.02$ &$14.03\pm0.02$  &--&--&--\\
12&  $16.33\pm0.05$   & $15.74\pm0.03$ &$14.87\pm0.02$ &$14.42\pm0.06$ &$14.00\pm0.05$ & $13.24\pm0.02$ &$12.82 \pm0.03  $&$12.74 \pm0.03 $\\ 
13&  $19.37\pm0.07$   & $18.77\pm0.01$ &$17.99\pm0.02$ &$17.52\pm0.03$ &$17.08\pm0.03$ & $16.35\pm0.13$ &$15.68 \pm0.13  $&$15.93 \pm0.29 $\\ 
14&  $16.83\pm0.04$   & $16.79\pm0.01$ &$16.24\pm0.03$ &$15.92\pm0.06$ &$15.57\pm0.02$ & $14.95\pm0.04$ &$14.73 \pm0.06  $&$14.79 \pm0.11 $\\ 
15&  $18.52\pm0.04$   & $17.84\pm0.01$ &$16.97\pm0.03$ &$16.46\pm0.03$ &$16.03\pm0.03$ & $15.26\pm0.05$ &$14.79 \pm0.07  $&$14.70 \pm0.10 $\\ 
\hline
\end{tabular} \\
 * No value for the uncertainty is given by the 2-MASS catalogue.
\label{sequence_magsJC}
\end{table*}

\begin{table*}
\centering
\caption{Sloan Vega magnitudes and associated errors for the stellar sequence used in the calibration process of SN~2011fu's photometry.}
\begin{tabular}{cccccc} \hline 
    Star    &  \textit{u}   & \textit{g*}  & \textit{r} & \textit{i}  & \textit{z} \\ 
            & (mag) & (mag) & (mag) & (mag)& (mag)  \\
  \hline
1 & --            & --     &$17.20\pm0.01$ &$16.64\pm0.01$ &$16.28\pm	0.01$ \\
2 & --            &$17.38\pm0.01$ &$17.15\pm0.01$ &$16.77\pm0.01$ &$16.52\pm	0.03$ \\
3 & --            &$17.44\pm0.01$ &$17.22\pm0.01$ &$16.83\pm0.01$ &$16.59\pm	0.01$ \\
4 & --            &$17.19\pm0.01$ &$16.63\pm0.01$ &$16.13\pm0.01$ &$15.82\pm	0.01$ \\
5 &$17.85\pm0.02$ &$17.04\pm0.01$ &$16.89\pm0.01$ &$16.54\pm0.01$ &$16.34\pm	0.02$ \\
6 &$15.52\pm0.03$ & --     &$14.45\pm0.01$ &$14.06\pm0.01$ &$13.83\pm	0.01$ \\
7 &$17.59\pm0.05$ &$16.96\pm0.01$ &$16.75\pm0.01$ &$16.36\pm0.01$ &$16.12\pm	0.01$ \\
8 & --            &$17.73\pm0.01$ &$16.83\pm0.01$ &$16.16\pm0.01$ &$15.71\pm	0.01$ \\
9 &$17.59\pm0.05$ &$16.88\pm0.01$ &$16.72\pm0.01$ &$16.38\pm0.01$ &$16.16\pm	0.02$ \\
10& --            &$17.96\pm0.01$ &$17.57\pm0.01$ &$17.12\pm0.01$ &$16.84\pm	0.03$ \\
11&$15.05\pm0.02$ & --     &$14.32\pm0.02$ &$13.98\pm0.02$ &$13.84\pm	0.04$ \\
12&$16.15\pm0.03$ & --     &$14.38\pm0.04$ &$13.97\pm0.01$ &$13.68\pm	0.01$ \\
13& --            &$17.87\pm0.01$ &$17.53\pm0.01$ &$17.05\pm0.01$ &$16.78\pm	0.02$ \\
14& --            & --     &$15.88\pm0.01$ &$15.52\pm0.01$ &$15.29\pm	0.01$ \\
15& --            &$15.95\pm0.01$ &$16.46\pm0.01$ &$16.00\pm0.01$ &$15.71\pm	0.01$ \\

\hline
\end{tabular} \\
* The stars were only measured in one epoch in this band.
\label{sequence_magssloan}
\end{table*}

\label{lastpage}
\end{document}